\documentclass[aps,prl,reprint,groupedaddress,showpacs,showkeys]{revtex4-1}
\pdfoutput=1 
 \usepackage{amsmath}
 \usepackage{amsfonts}
 \usepackage{graphicx}
 \usepackage{epsfig}
 \usepackage{epstopdf}
 \usepackage{newtxtext,newtxmath}
 \usepackage{comment}

 \usepackage{hyperref}
 \hypersetup{
    bookmarks=true,         
    unicode=false,          
    pdftoolbar=true,        
    pdfmenubar=true,        
    pdffitwindow=false,     
    pdfstartview={FitH},    
    pdftitle={Many-body theory of positronium-atom scattering},   
    pdfauthor={D. G. Green},     
    pdfsubject={PsAtomMBT},   
    pdfcreator={D. G. Green},   
    pdfproducer={Producer}, 
    pdfkeywords={keywords}, 
    pdfnewwindow=true,      
    colorlinks=true,       
    linkcolor=blue,          
    citecolor=blue,        
    filecolor=magenta,      
    urlcolor=blue           
}

\graphicspath{{figs/}}

\begin{document}

\frenchspacing

\title{Many-body theory for positronium-atom interactions}
\author{D.~G. Green}
\email{d.green@qub.ac.uk}
\author{A.~R. Swann}
\email{a.swann@qub.ac.uk}
\author{G.~F. Gribakin}
\email{g.gribakin@qub.ac.uk}
\affiliation{
Centre for Theoretical Atomic, Molecular, and Optical Physics, School of Mathematics and Physics,
Queen's University Belfast, Belfast BT7\,1NN, Northern Ireland, United Kingdom}
\date{\today}

\begin{abstract}
A many-body-theory approach has been developed to study positronium-atom interactions.
As first applications, we calculate the elastic scattering and momentum-transfer cross sections and the pickoff annihilation rate $^1Z_\text{eff}$ for Ps collisions with He and Ne. For He the cross section is in agreement with previous coupled-state calculations, while comparison with experiment for both atoms highlights discrepancies between various sets of measured data. In contrast, the calculated $^1Z_\text{eff}$ (0.13 and 0.26 for He and Ne, respectively) are in excellent agreement with the measured values.
\end{abstract}

\maketitle

Positronium (Ps) is a light ``atom'' consisting of an electron and its antiparticle, the positron. It is important for precision tests of QED \cite{Karshenboim2005} and for understanding galactic positron annihilation \cite{MNR:MNR17804}. It also has numerous applications, from probing free space in condensed matter systems \cite{Gidley06} to making antihydrogen \cite{Kellerbauer08} and studying free fall of antimatter \cite{PhysRevLett.106.173401}. These applications require understanding of Ps interaction with normal matter, which is far from complete. Recent experiments on Ps scattering on noble-gas atoms revealed some unexpected trends, e.g., that the scattering cross section becomes very small at low Ps energies \cite{Brawley:2015}. Overall, there is a large uncertainty in the existing Ps-atom scattering data \cite{Swann:2018:vdw}, while calculations of the rate of \textit{pickoff} annihilation in noble gases (where the positron from Ps annihilates with an atomic electron) \cite{Fraser:1966,Fraser:1968,Barker:1968,Drachman:1970,Biswas:2000,Mitroy:2001,Mitroy:2003} underestimate the experimental data \cite{Charlton:1985,Saito:2006} by as much as a factor of ten.

The theoretical description of Ps-atom interactions is challenging because of the composite nature of the collision partners and a significant cancellation between the short-range Ps-atom repulsion and van-der-Waals attraction. Accurate calculations must account for dynamical distortion of both objects during the collision, which has only been achieved for simple targets, i.e., hydrogen and helium \cite{Walters:2004}. Calculations of pickoff annihilation require  account of important short-range electron-positron correlations, which provide corrections to the annihilation vertex \cite{Dunlop:2006,DGG_posnobles,DGG:2015:core,DGG_hlike}, but have been neglected in all previous calculations \footnote{The only exception is a calculation for Ps-H$_2$
that uses explicitly correlated Gaussians \cite{Zhang:2018}.}.

Many-body theory (MBT) is a powerful and systematic method of accounting for virtual excitations of both objects and the electron-positron correlation effects. It provided an accurate description of low-energy electron-atom scattering \cite{Kelly:MBT:elcAtom,Amusia:elc:MBT:Ar1,Amusia:JETP:1975,Amusia:elc:MBT:Ar2,Johnson:MBT:elc:Xe,Safronova:MBT:elcAtom} and positron interaction with atoms \cite{Amusia:Pos:MBT:He,PhysScripta.46.248,Gribakin:2004,DGG_posnobles,DGG:2015:core,DGG_posnobles,DGG:2015:core,DGG_cool,DGG_gamcool}, with scattering cross sections, annihilation rates, and $\gamma$ spectra all found to be in excellent agreement with experiment. 

In this Letter we show how to describe Ps interaction with a many-electron atom by combining the MBT description of electron-atom and positron-atom interactions, and including the important effect of screening of the electron-positron Coulomb interaction by the atom.
As first applications of the theory, we calculate phase shifts, elastic-scattering and momentum-transfer cross sections, and the pickoff annihilation rate $^1Z_\text{eff}$ for Ps on He and Ne.
The cross sections are found to be in agreement with previous coupled-state \cite{Walters:2004} and  model van der Waals \cite{Swann:2018:vdw} calculations. 
By accounting for electron-positron correlation corrections to the annihilation vertex, we obtain values of $^1Z_\text{eff}$ in excellent agreement with experiment \cite{Charlton:1985}.
Atomic units (a.u.) are used throughout.

\textit{MBT of electron- and positron-atom interactions.}---MBT describes an electron or positron in the field of a many-electron atom via the Dyson equation for the (quasiparticle) wave function $\psi_\varepsilon$ \cite{fetterwalecka}:
\begin{equation}\label{eqn:dyson}
\bigl(\hat{H}^{\pm}_0+\hat{\Sigma}^{\pm}_{\varepsilon}\bigr)\psi^\pm_{\varepsilon}({\bf r})=\varepsilon\psi^\pm_{\varepsilon}({\bf r}).
\end{equation}
Here $\hat{H}^{\pm}_0$ is the zeroth-order Hamiltonian, e.g., that of the electron $(-)$ or positron $(+)$ in the field of the Hartree-Fock (HF) ground-state atom, and $\hat{\Sigma}^{\pm}_{\varepsilon}$ is the nonlocal, energy-dependent correlation potential  \footnote{$\hat{\Sigma}^{\pm}_{E}$ acts as $\hat{\Sigma}^\pm_E\psi_{\varepsilon}({\bf r}) = \int \Sigma_E^\pm({\bf r},{\bf r}')\psi_{\varepsilon}({\bf r'})\, d{\bf r}'$. }, equal to the electron or positron self-energy in the field of the atom.
Equation~(\ref{eqn:dyson}) can be solved separately for each partial wave, with the wave function in the form $\psi^\pm_{\varepsilon}({\bf r}) = r^{-1}\tilde{P}^\pm_{\varepsilon \ell}(r)Y_{\ell m}(\hat{\bf r})$, where $Y_{\ell m}$ is a spherical harmonic. 
\begin{figure}[t!!]
\includegraphics*[width=0.48\textwidth]{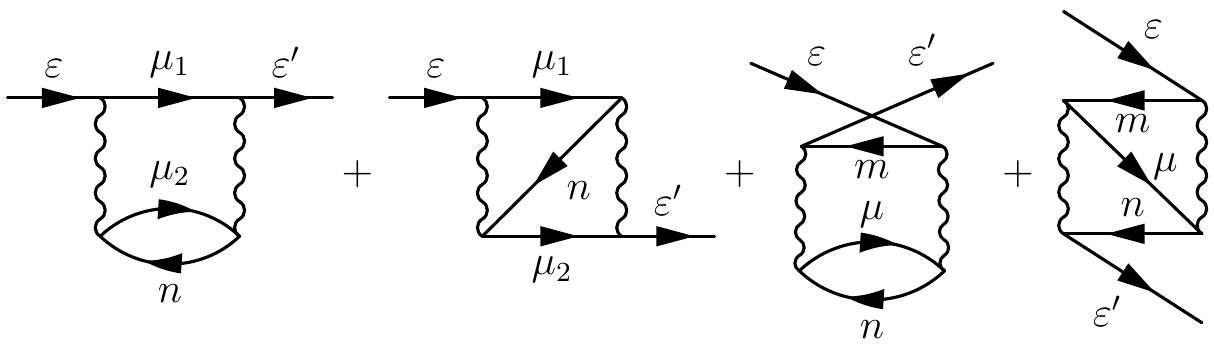}
\includegraphics*[width=0.36\textwidth]{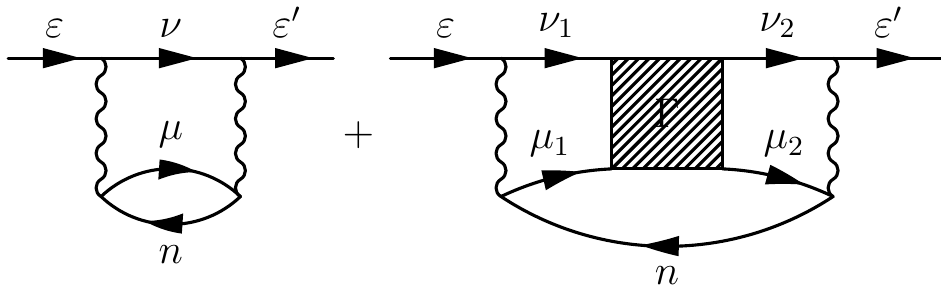}
\caption{The main contributions to the self-energy of the electron (top line) and positron (bottom line) in the field of the atom. 
Lines labeled $\varepsilon$ represent the electron or positron HF wave function. Lines labeled $\nu$ ($\mu$) represent positron (excited electron) states, which are summed over. Lines labeled $n$ and $m$ represent holes in the atomic ground state.  Wavy lines represent  Coulomb interactions.
The shaded $\Gamma$ block represents the sum of the electron-positron ladder-diagram series \cite{Gribakin:2004,DGG_posnobles}, which accounts for virtual Ps formation.
\label{fig:selfnrg}}
\end{figure}
Rather than computing the self-energy $\Sigma_E^\pm({{\bf r,r}'})$ in coordinate space, it is more convenient to work with its matrix elements $\langle \varepsilon' |\hat \Sigma^{\pm}_E| \varepsilon \rangle$ in the HF basis $\left\{\varphi^\pm_{\varepsilon}\right\}$, where
$H_0^\pm\varphi^\pm_{\varepsilon}=\varepsilon\varphi^\pm_{\varepsilon}$, 
$\varphi^\pm_{\varepsilon}({\bf r})=r^{-1}P^\pm_{\varepsilon \ell}(r)Y_{\ell m}(\hat{\bf r})$, and
$\langle \varepsilon' |\hat \Sigma^{\pm}_E| \varepsilon \rangle =
\int P^\pm_{\varepsilon' \ell}(r')\Sigma_{E\ell}^{\pm}(r,r') P^\pm_{\varepsilon l}(r) \,dr\, dr'$,
with $\Sigma_{E\ell}^\pm$  the self-energy for partial wave $\ell$. Using the completeness of the basis, it can be expressed as
\begin{eqnarray}\label{eqn:sigmarr}
\Sigma^{\pm}_{E \ell}(r,r') = \sum_{\varepsilon,\varepsilon'} P^\pm_{\varepsilon' \ell}(r') 
\langle \varepsilon'
|\Sigma^{\pm}_{E\ell} | 
\varepsilon \rangle P^\pm_{\varepsilon \ell}(r).
\end{eqnarray}

Figure~\ref{fig:selfnrg} shows the main contributions to the electron and positron self-energy. For the electron (top row), the first diagram accounts for the attractive long-range polarization potential $-\alpha_d/2r^{4}$, where $\alpha_d$ is the dipole polarizability of the atom. The other three diagrams contribute only at short range. These diagrams provide a good description of the electron interaction with noble-gas atoms \cite{Amusia:elc:MBT:Ar1,Amusia:JETP:1975,Amusia:elc:MBT:Ar2, Johnson:MBT:elc:Xe}\footnote{See Ref.~\cite{Safronova:MBT:elcAtom} for higher-order calculations.}.
For the positron (bottom row), the first diagram produces a long-range polarization potential similar to that for the electron. The second diagram describes an important contribution of virtual Ps formation \cite{PhysScripta.46.248,Gribakin:2004,DGG_posnobles}. Here the $\Gamma$ block represents the sum of the infinite electron-positron ladder-diagram series \cite{Gribakin:2004}. We calculate the electron and positron self-energies as described in Ref.~\cite{DGG_posnobles}, using a $B$-spline basis with 40 splines of order 6 defined over an exponential knot sequence, discretizing the continuum by confining the system in a spherical cavity of radius 30 a.u. The corresponding electron and positron basis sets ensure convergence of the sums over intermediate states.

\begin{figure}[t!!]
\centering
\includegraphics[width=0.48\textwidth]{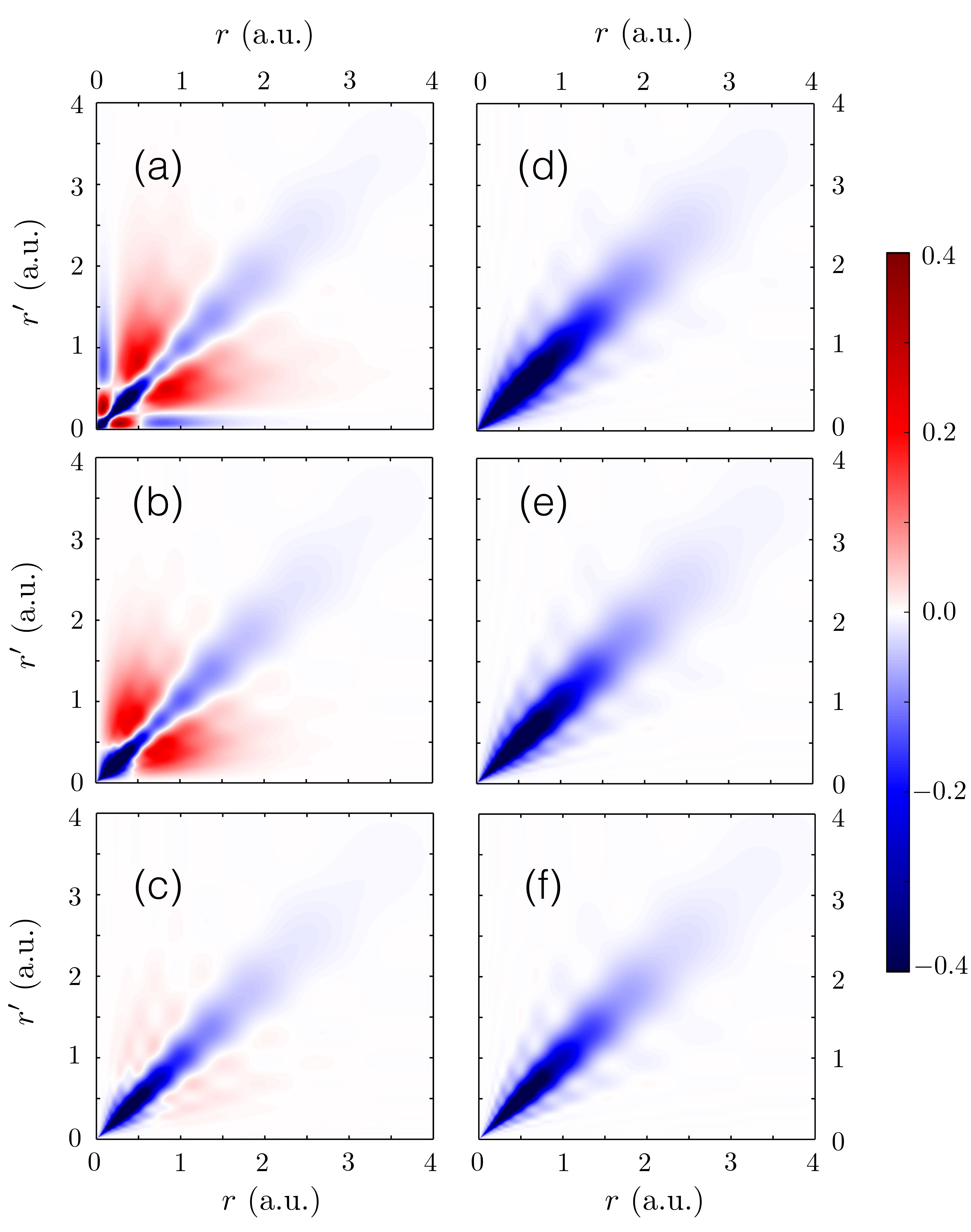}\\
\caption{Self-energy  $\Sigma^\pm_{0\ell}(r,r')$ for Ne for the electron (a) $\ell = 0$, (b) $\ell = 1$, and (c) $\ell = 2$; and positron (d) $\ell = 0$, (e) $\ell = 1$, and (f) $\ell = 2$. \label{fig:sigr}}
\end{figure}

The correlation potential described above is essentially nonlocal. It is also quite different for the electron and positron, and for different partial waves. Figure~\ref{fig:sigr} shows $\Sigma^{\pm}_{E\ell}(r,r')$ for the $s$, $p$, and $d$ waves in Ne, calculated at $E=0$. 
Their key feature is a ``valley'' along the diagonal $r = r'$, whose width characterizes the degree to which $\hat \Sigma^{\pm}$ is nonlocal. The main contribution to electron- and positron-atom attraction comes from $r\gtrsim 1$ a.u.~(i.e., outside the atom). Here, $\Sigma^+_{E\ell}(r,r')$ is more negative than $\Sigma^-_{E\ell}(r,r')$, meaning stronger attraction for the positron.
As a consequence of the Pauli principle, the correlation potential for the electron is quite different for different partial waves. It is also significantly more nonlocal than that of the positron, with prominent repulsive areas for the $s$ and $p$ waves. These features are due to the contribution of the second, exchange diagram to $\Sigma^{-}_{E\ell}(r,r')$.  

\begin{figure}[t!!]
\includegraphics[width=0.3\textwidth]{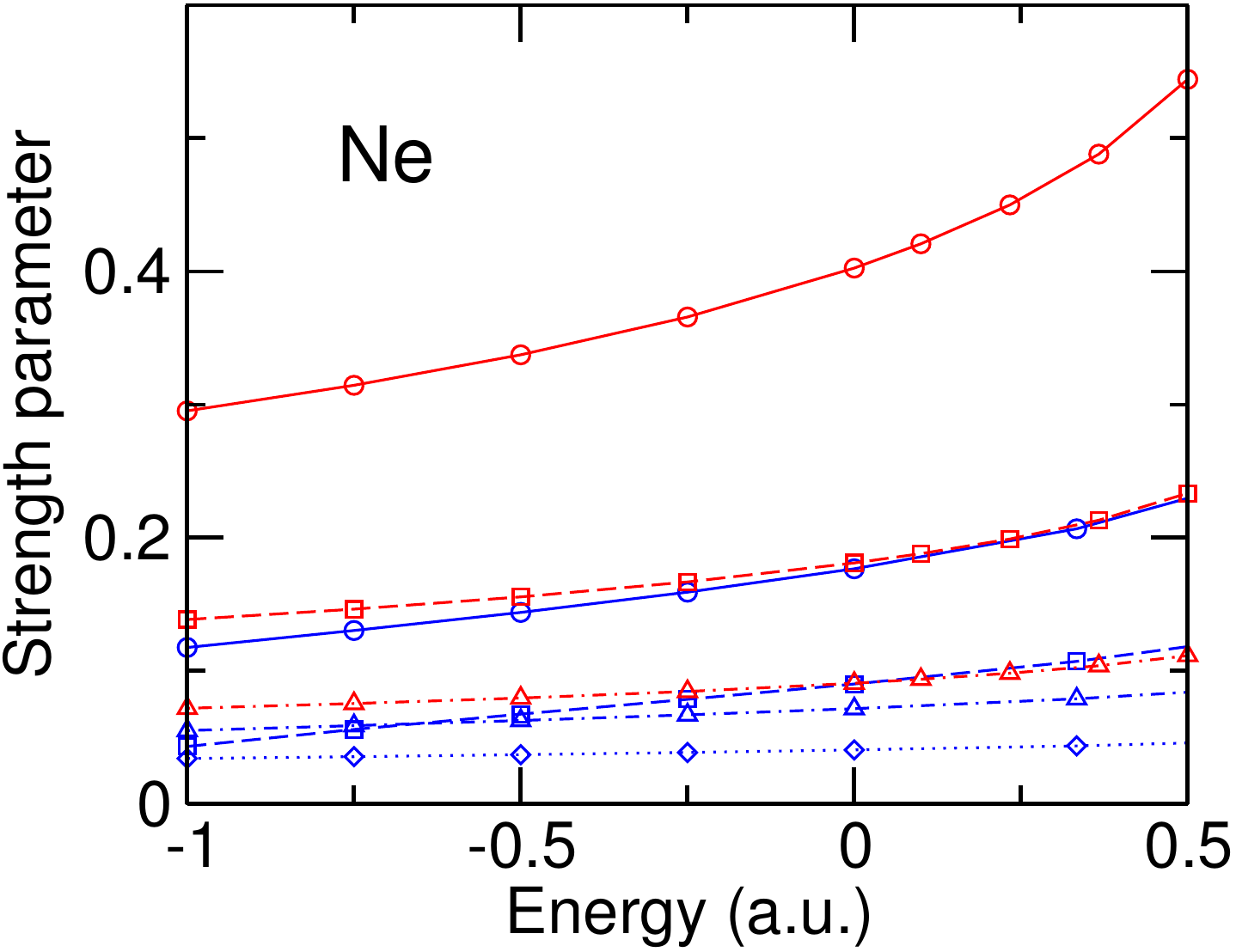}
\caption{The dimensionless strength parameter $\mathcal{S}^{\pm}_{E\ell}$ 
of the correlation potentials for the electron (blue) and positron (red) in the field of Ne, as a function of energy $E$,  for $\ell=0$ (circles), $\ell=1$ (squares), $\ell=2$ (triangles), and $\ell=3$ (diamonds).\label{fig:S}}
\end{figure}

The energy dependence of the electron and positron correlation potentials can be analyzed by examining the dimensionless strength parameter $\mathcal{S}^{\pm}_{E\ell} = -\sum_{\varepsilon>0}\langle \varepsilon | \Sigma^{\pm}_{E\ell} | \varepsilon\rangle/ \varepsilon$ \cite{Dzuba:1994}.
Figure~\ref{fig:S} shows $\mathcal{S}^{\pm}_{E\ell}$ for Ne, as a function of energy for electron and positron $s$, $p$, $d$, and (for the electron) $f$  waves.
It confirms that  the correlation potential is stronger for the positron. It also shows that its energy dependence is relatively weak on the energy scale of Ps (0.25~a.u.). This is important for the description of Ps-atom interaction, as it allows us to use $\Sigma^\pm_{E\ell}(r,r')$
calculated for a fixed energy ($E=0$).

\begin{figure}[b!!]
\includegraphics[width=0.48\textwidth]{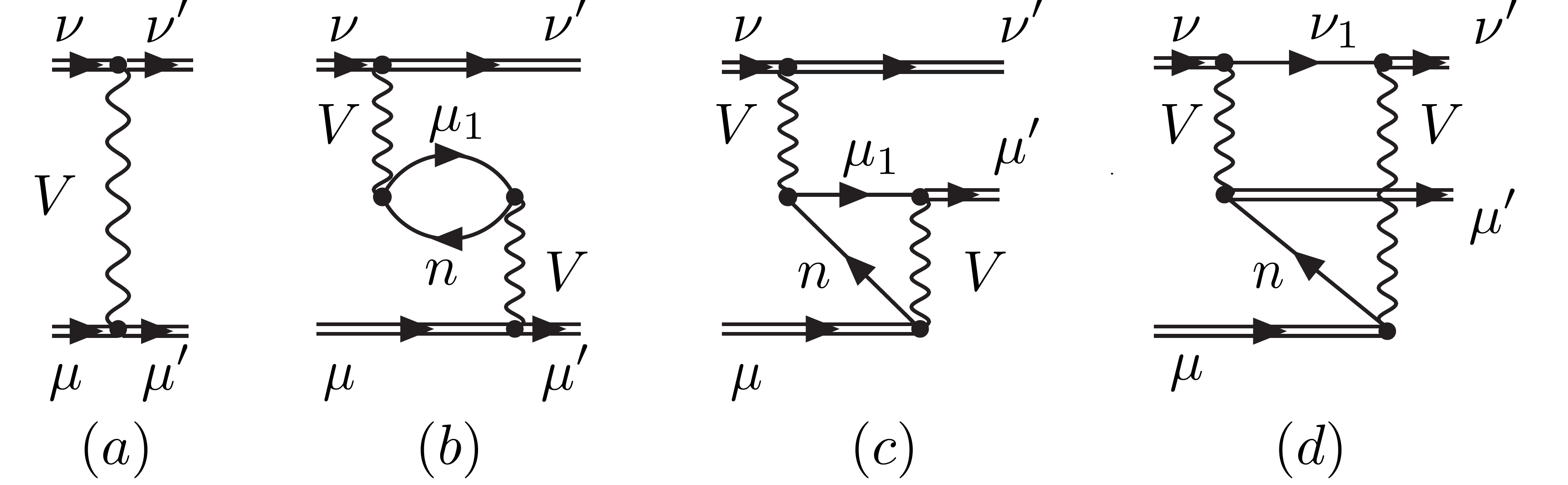}
\caption{The main contributions to the electron-positron interaction in Ps: (a) the bare Coulomb interaction $V$; (b)--(d), screening $\delta V_E$ with exchange contributions (with mirror images). Double lines labeled $\nu$ ($\mu$) represent positron (electron) Dyson states in the field of the atom.
\label{fig:tildev}}
\end{figure}

\textit{MBT of Ps-atom interactions.}---
The wave function $\Psi$ of Ps in the field of the atom satisfies the two-particle Dyson equation (also known as the Bethe-Salpeter equation \cite{fetterwalecka})
\begin{eqnarray}\label{eqn:tpde}
\bigl(
\hat{H}_0^{-}+\hat{\Sigma}^-_{\varepsilon ^-}+
\hat{H}_0^{+}+\hat{\Sigma}^+_{\varepsilon ^+} 
+ V+\delta V_E\bigr) \Psi=E\Psi ,
\end{eqnarray}
where $V$ is the electron-positron Coulomb interaction and $\delta V_E$ is the screening correction due to polarization of the atom \footnote{There is a similarity between our approach and the combination of MBT with the configuration-interaction method for open-shell atoms \cite{Dzuba:MBTCI}}.
The diagrams for $\delta V_E$ are shown in Fig.~\ref{fig:tildev}.
The main screening diagram Fig.~\ref{fig:tildev} (b) is essential for canceling the long-range $r^{-4}$ polarization attraction and making the long-range Ps-atom interaction of the required $R^{-6}$ van der Waals form, where $R$ is the distance between the Ps center of mass and the atom.
The exchange corrections Fig.~\ref{fig:tildev} (c) and (d) are typically much smaller. They also partly cancel each other and can be neglected.

We construct the Ps eigenstates with angular momentum $J$ and parity $\Pi$ from the single-particle Dyson states \cite{Brown:2017,Swann:2018:vdw}, as
\begin{eqnarray}\label{eqn:ps}
\Psi_{J\Pi}({\bf r}_e,{\bf r}_p) = \sum_{\mu,\nu} C^{J\Pi}_{\mu\nu} \psi^-_{\mu}({\bf r}_e)\psi^+_{\nu}({\bf r}_p).
\end{eqnarray}
The energy eigenvalues $E$ and coefficients $C^{J\Pi}_{\mu\nu}$ are found by solving matrix eigenvalue problem for the Hamiltonian matrix
\begin{eqnarray}\label{eqn:ham}
\langle\nu'\mu'|H|\mu\nu\rangle=
\left(\varepsilon_{\mu}+\varepsilon_{\nu}\right)\delta_{\mu'\mu}\delta_{\nu'\nu} + \langle\nu'\mu'|V+\delta V_E|\mu\nu\rangle.
\end{eqnarray}
We consider $J^\Pi=0^+$, $1^-$, and $2^+$ to investigate Ps $S$-, $P$-, and $D$-wave scattering, respectively. To ensure accurate description of Ps states by Eq.~(\ref{eqn:ps}), we confine the electron and positron states to a cavity of radius $R_c=10$--16~a.u. \cite{Brown:2017}. To represent the positive-energy ``continuum'' in the cavity, we use a second $B$-spline basis of 60 splines of order 9 defined over a quadratic-linear knot sequence \cite{Swann:2018:vdw}. The effect of $\Sigma^{\pm}_{E\ell}$ decreases with $\ell$, and we find that it sufficient to use  Dyson  states in Eq.~(\ref{eqn:ps}) for $\ell\leq3$, and HF states for higher $\ell$. 
We exploit the weak energy dependence of $\Sigma_{E}^{\pm}$ and $\delta V_E$ 
by evaluating them at $E=0$.
Calculations are performed with different numbers of radial states and angular momenta included in Eq.~(\ref{eqn:ps}), up to $n_{\rm max}=20$ and $\ell_{\rm max}=20$. Such high angular momenta are required to ensure convergence of the Ps wave function, which is given by a single-centre expansion about the atomic nucleus. Accurate Ps states are found by extrapolating to $n_{\rm max}\to\infty$ and $\ell_{\rm max}\to\infty$ (see Ref. \cite{Brown:2017} for details).

\emph{Ps scattering on He and Ne}.---As a first application, we calculate the phase shifts and cross sections for Ps scattering on He and Ne. The phase shifts are determined from the Ps energy eigenvalues, as described in Ref.~\cite{Swann:2018:vdw}. Calculations were performed using cavity radii of $10$, 12, 14, and 16~a.u. Effective-range-type fits were used to interpolate the $S$, $P$, and $D$ phase shifts calculated at the discrete values of the Ps center-of-mass momentum $K$. The phase shifts yield values the scattering length and the partial contributions to the elastic and momentum-transfer cross sections.

\begin{figure}[t!!]
\includegraphics[width=4.4cm]{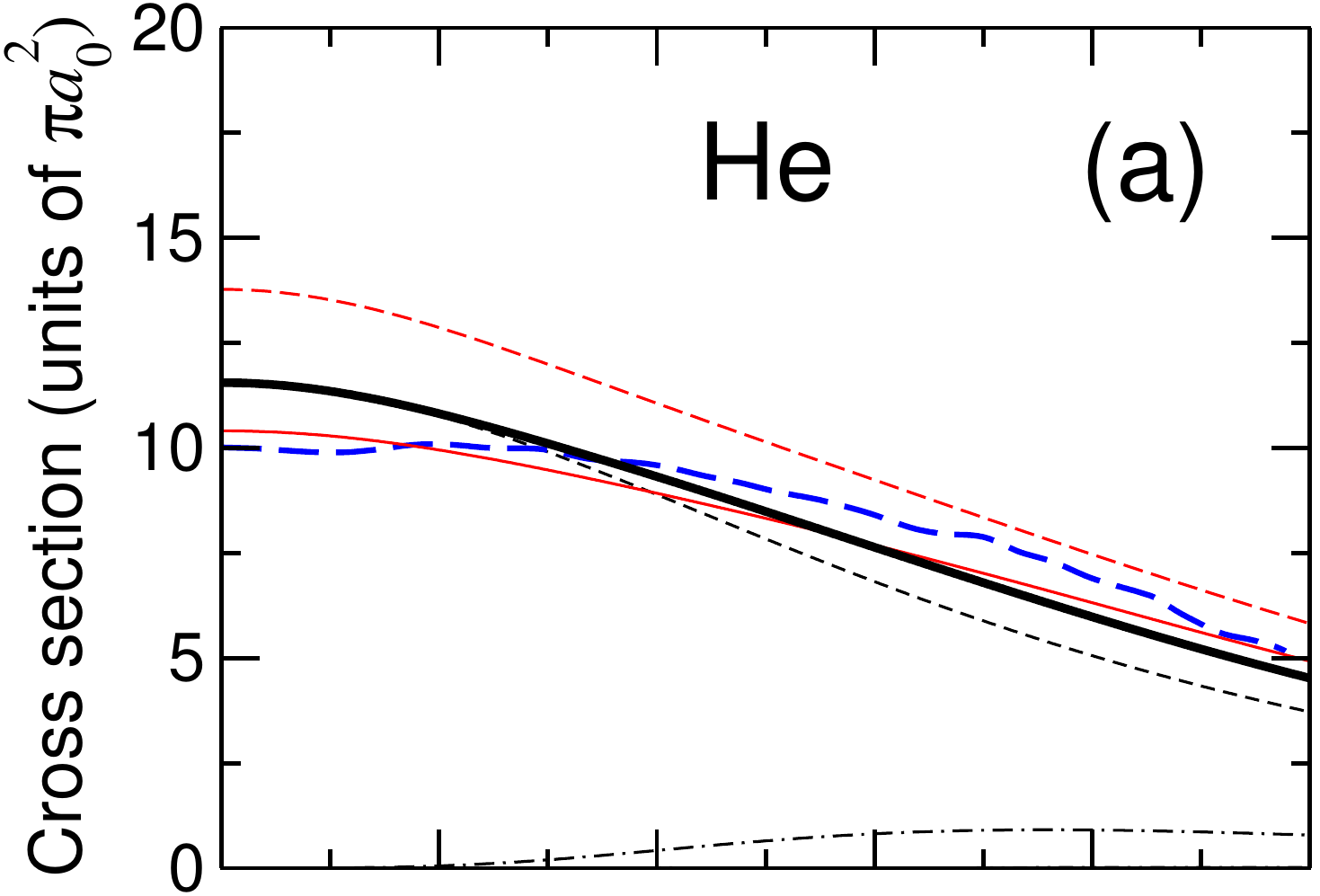}
\includegraphics[width=4cm]{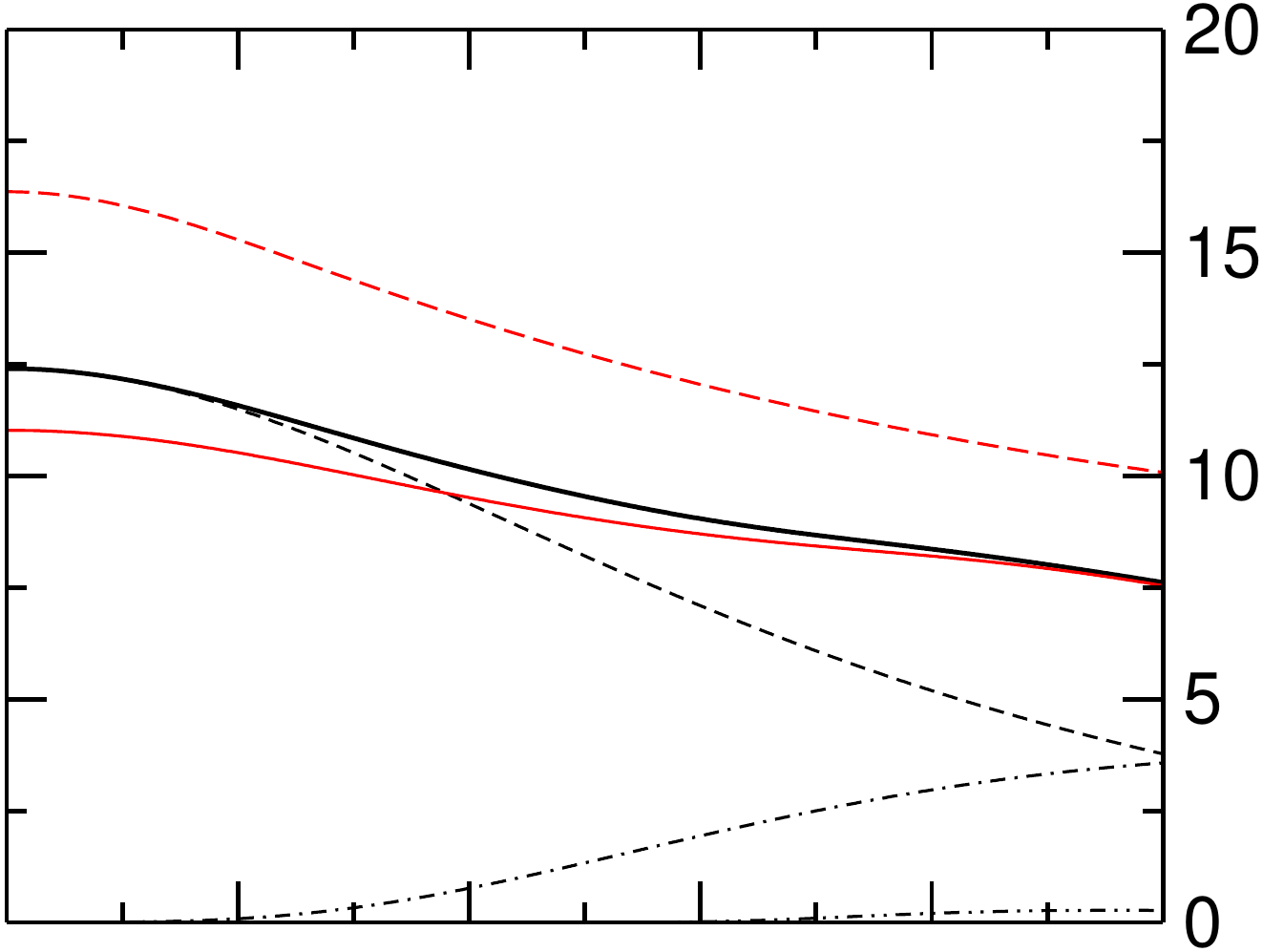}

\includegraphics[width=4.4cm]{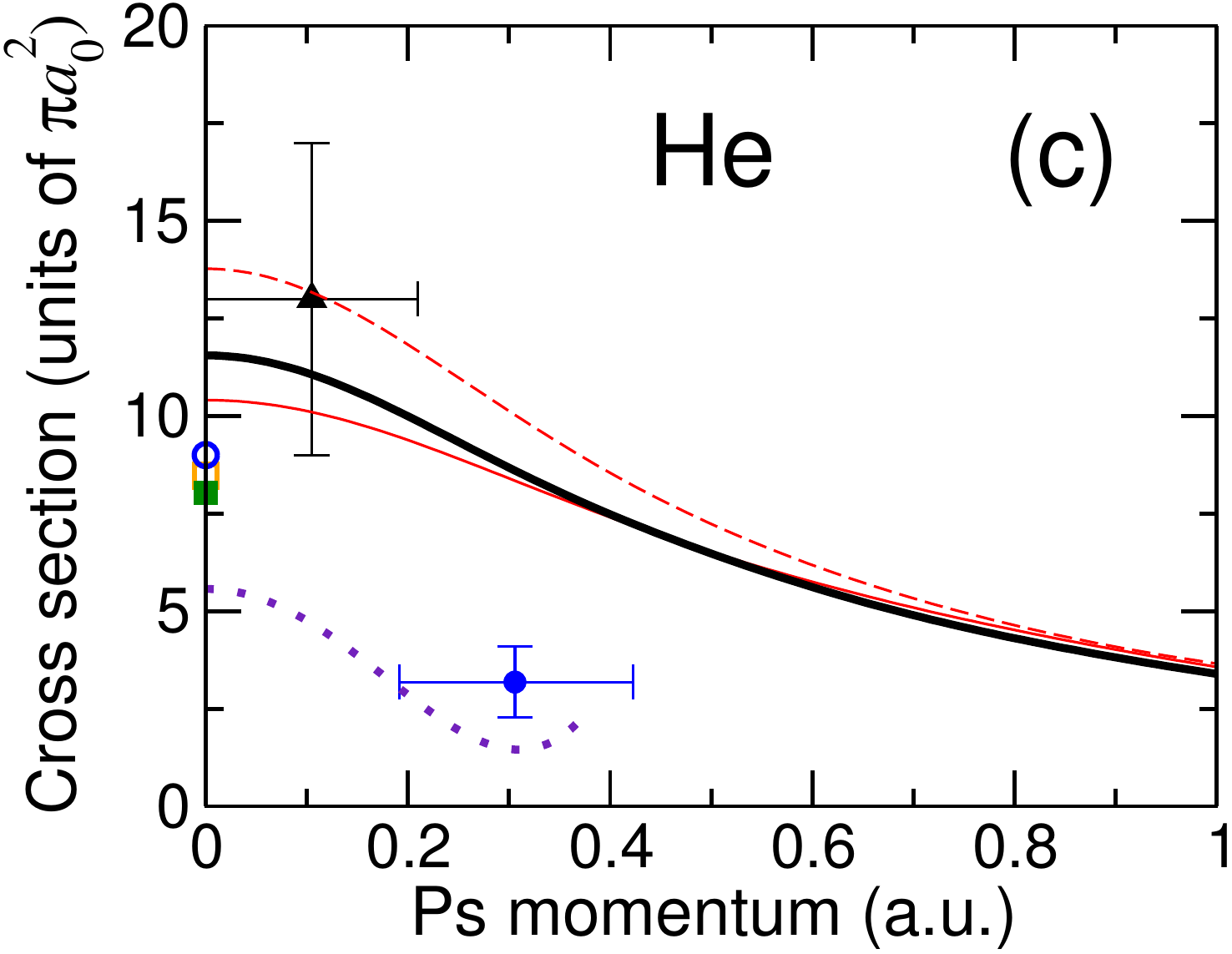}
\includegraphics[width=4cm]{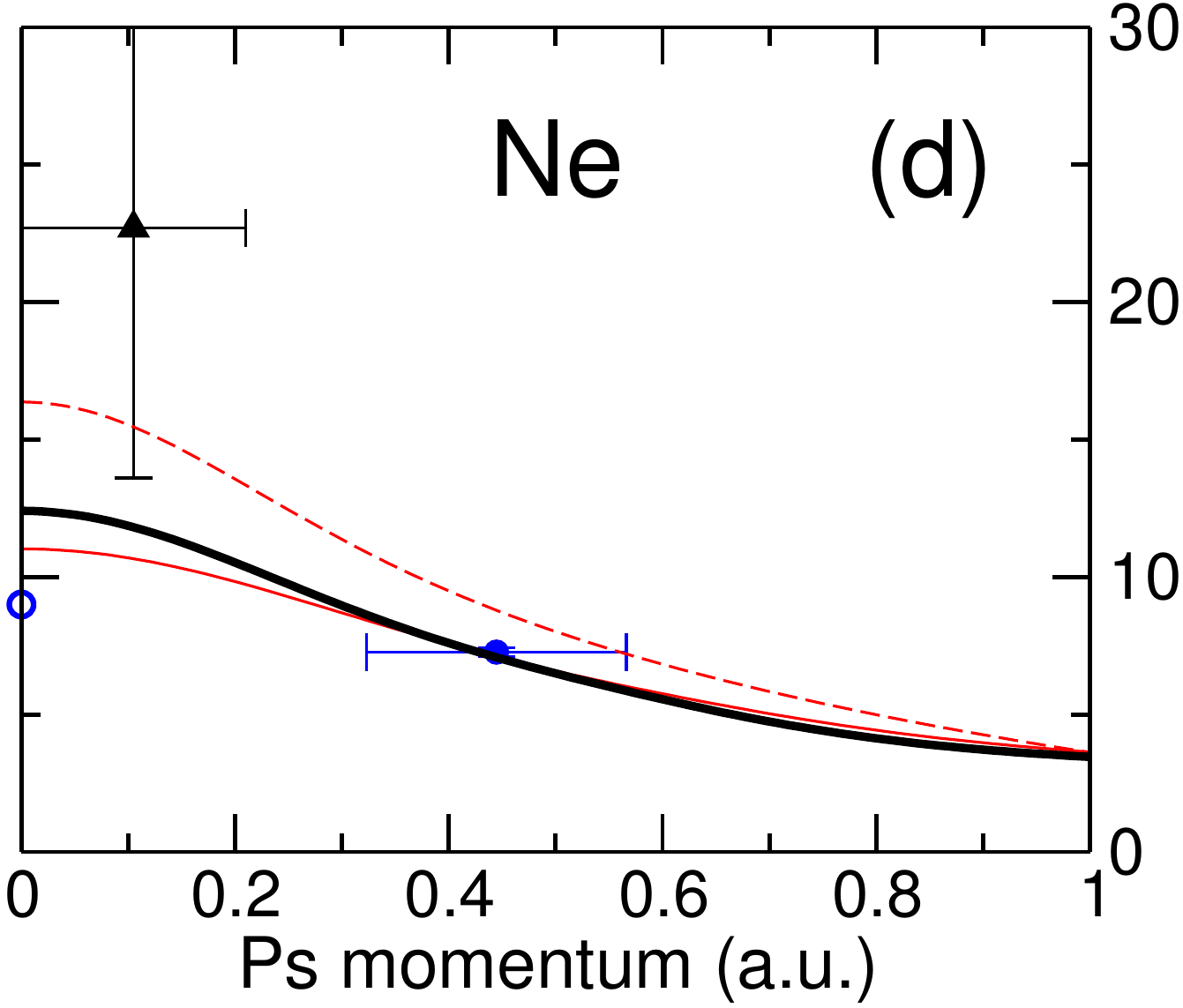}
\caption{Elastic-scattering cross sections for Ps on He (a) and Ne (b):
total cross section calculated using MBT (thick solid line), with $S$- (dashed black line), $P$- (dot-dashed black line), and $D$-wave (dot-dot-dashed black line) partial contributions;
FT (dashed red line) and FT+vdW (thin solid red line) calculations of Ref.~\cite{Swann:2018:vdw}. Also shown for He is the 9-Ps-9-He-state calculation of Walters \emph{et al.}~\cite{Walters:2004} (thick dashed blue line).
Momentum-transfer cross sections for Ps on He (c) and Ne (d) use the same symbols as in (a) and (b). Also shown are the experimental results
\cite{Canter:1975} (open square),
\cite{Rytsola:1984} (filled square),
\cite{Coleman:1994} (open circle),
\cite{Skalsey:2003} (filled circle),
\cite{Nagashima:1998,Saito:2003} (triangle), and
\cite{Engbrecht:2008} (dotted line).
\label{fig:xsecs}}
\end{figure}

The  partial and total elastic scattering cross sections are shown in Fig.~\ref{fig:xsecs} (a) and (b) for He and Ne, respectively.
Comparing with the frozen-target (FT) results [obtained by neglecting $\hat \Sigma ^\pm _\varepsilon $ and $\delta V_E$ in Eq.~(\ref{eqn:tpde})], we see that correlations partially cancel the FT Ps-atom repulsion and reduce the cross sections.
For He, the $S$-wave contribution dominates across the range of momenta considered, but for Ne, the $P$-wave contribution becomes comparable at $K\approx1$ a.u.
For He, the elastic cross section is close to the 9-Ps-9-He coupled-state calculation of Walters \emph{et al.}~\cite{Walters:2004}, and to the calculation \cite{Swann:2018:vdw} in which a model van-der-Waals potential was added to the FT Ps-atom interaction (FT+vdW) \footnote{The FT+vdW data corresponds to the cutoff radius $R_0=3.0$~a.u. in the model van der Waals potential \cite{Swann:2018:vdw}.}. The MBT scattering length of 1.70 a.u. compares well with the value of 1.6~a.u. obtained in Ref.~\cite{Walters:2004}. It is ${\sim} 10\%$ smaller than the FT value (1.86 a.u. \cite{Swann:2018:vdw}), highlighting the importance of including distortion of the target.
For Ne, the MBT scattering length of 1.76 a.u. is ${\sim}$15\% smaller than the FT value (2.02 a.u.) but close to the FT+vdW result (1.66 a.u. \cite{Swann:2018:vdw}). The relatively small effect of the correlations, i.e., the difference between the MBT and FT calculations, is due to cancelation between the positron- and electron-atom attraction ($\hat \Sigma ^\pm _\varepsilon $) and the effect of screening $\delta V_E$. It is worth noting that while the phase shifts and cross sections from MBT and FT+vdW calculations are close, they cannot be reproduced by a simple local potential, such as of Lennard-Jones form \footnote{We have checked that a Ps-atom potential of the form $U(R)=B/R^{12}-C_6/R^6$ (as suggested in Ref.~\cite{Nieminen:1980}) with accurate $C_6$ \cite{Swann:2015} and $B$ chosen to fit the MBT scattering length, gives the elastic cross sections that exceed the MBT result by 50\% (Ne) to 100\% (He) at $K\sim 1$~a.u., due to overestimated effect of repulsion.}.

The MBT results for the momentum-transfer cross section [Fig.~\ref{fig:xsecs} (c) and (d)], are close to the FT+vdW calculation \cite{Swann:2018:vdw}, particularly for $K>0.3$~a.u. For He, our calculation is within the error bars of the experimental result of Nagashima \emph{et al.}~\cite{Nagashima:1998} but $\sim$30--45\% larger than that of Canter \emph{et al.}~\cite{Canter:1975}, Rytsola \emph{et al.}~\cite{Rytsola:1984}, and Coleman \emph{et al.}~\cite{Coleman:1994}. The measurements of Skalsey \emph{et al.}~\cite{Skalsey:2003} and Engbrecht \emph{et al.}~\cite{Engbrecht:2008} give much lower values. These measurements are based on Doppler-broadening spectroscopy (DBS), and may suffer from errors related to the discrimination of the narrow Ps annihilation component on the background of the positron-He annihilation signal. This background is much broader for Ne, which is
possibly explains why the DBS data from Skalsey \emph{et al.}~\cite{Skalsey:2003} are in good agreement with the MBT results. At the same time, the MBT result for Ne
is just outside the error bars of Saito \emph{et al.}~\cite{Saito:2003} and ${\sim}40\%$ greater than that of Coleman \emph{et al.}~\cite{Coleman:1994}.

\textit{Calculation of pickoff annihilation rates.}---The Ps pickoff annihilation rate in a gas is parametrized as $\lambda=4\pi r_0^2 c n_g {^1Z_\mathrm{eff}}$, where $r_0$ is the classical electron radius, $c$ is the speed of light, $n_g$ is the number density of the gas, and $^1Z_\mathrm{eff}$ is the effective number of electrons per atom in a singlet state relative to the positron \cite{Fraser:1966}. Our interest is in $^1Z_\mathrm{eff}$ values at small (thermal) Ps momenta, where only the $S$ wave contributes. In the zeroth-order, independent-particle approximation (IPA), it is given by
\begin{eqnarray}\label{eqn:zeff}
{^1Z}_{\rm eff}^{(0)} = \frac{1}{4}\sum_n \iint |\Psi_{0^+}({\bf r}_e,{\bf r}_p)|^2 |\varphi_n({\bf r}_p)|^2\, d{\bf r}_e\, d{\bf r}_p,
\end{eqnarray}
where the sum is over all HF orbitals $\varphi_n$ occupied in the ground-state atom, and $\Psi_{0^+}$ is normalized to a plane wave of the Ps center-of-mass motion far from the atom.
Previous IPA calculations for He \cite{Fraser:1966,Fraser:1968,Barker:1968,Barker:1969,Drachman:1970,Biswas:2000,Mitroy:2001,Swann:2018:vdw} and Ne \cite{Mitroy:2001,Swann:2018:vdw} yielded values of ${^1Z}_{\rm eff}^{(0)}$ that underestimated experimental data by a factor of 3 or more (see Table \ref{tab:zeff}).
\begin{table}
\caption{Pickoff annihilation rates $^1Z_\text{eff}$ for He and Ne at $K=0$: best previous theory \cite{Mitroy:2001}; using frozen-target  Ps wave function from Ref.~\cite{Swann:2018:vdw}; present theory, zeroth-order approximation (MBT); present theory with enhancement factors (MBT-EF); and experiment \cite{Charlton:1985}.}
\centering
\begin{ruledtabular}
\begin{tabular}{cccccc}
Atom & Ref.~\cite{Mitroy:2001} & FT \cite{Swann:2018:vdw} & MBT & MBT-EF & Exp. \cite{Charlton:1985}\\
\hline
He	& 0.0378 	& 0.0273 	& 0.0411 & 0.131 & 0.125\\
Ne	& 0.0922  	&0.0512	& 0.0932 & 0.255 & 0.235\\
\end{tabular}
\end{ruledtabular}
\label{tab:zeff}
\end{table}%
These calculations neglected the short-range electron-positron correlations, which are known to enhance the annihilation rates by a factor 2--5 \cite{DGG:2015:core,DGG:2017:ef}.

We account for the correlation corrections in ${^1Z}_{\rm eff}$ by augmenting Eq.~(\ref{eqn:zeff}) with \emph{enhancement factors} $\gamma_{n\ell}$, which are specific to the electron orbital $n$ and positron partial wave $\ell$ and were calculated in Refs.~\cite{DGG:2015:core,DGG:2017:ef}.
Explicitly, substituting Eq.~(\ref{eqn:ps}) into Eq.~(\ref{eqn:zeff}), and introducing the enhancement factors, yields
\begin{equation}
^1Z_\text{eff} = \frac14 \sum_{n,\mu,\nu , \nu'} \gamma_{n\ell} C^{0^+}_{\mu\nu} {C^{0^+}_{\mu\nu'}}^* \int \psi^+_\nu(\mathbf r) \left[\psi^+_{\nu'}(\mathbf r)\right]^* |\varphi_n({\bf r})|^2 \, d\mathbf r ,
\end{equation}
where the positron basis states $\psi^+_\nu$ and $\psi^+_{\nu'}$ both have angular momentum $\ell$.
Table~\ref{tab:enhancement_factors} shows the values of $\gamma_{n\ell}$ used.
\begin{table}
\caption{\label{tab:enhancement_factors}Enhancement factors $\gamma_{n\ell}$ for electron orbital $n$ and positron partial wave $\ell$, as calculated in Ref.~\cite{DGG:2017:ef}.}
\centering
\begin{ruledtabular}
\begin{tabular}{ccccc}
Atom & $n$ & $\ell =0$ & $\ell =1$ & $\ell =2$ \\
\hline
He & $1s$ & 2.99 & 4.04 & 5.26 \\
Ne & $1s$ & 1.18 & 1.21 & 1.22 \\
Ne &$2s$ & 1.87 & 2.03 & 2.30 \\
Ne &$2p$ & 2.78 & 3.46 & 4.70 
\end{tabular}
\end{ruledtabular}
\label{default}
\end{table}%

We perform calculations for the lowest-energy $J^{\Pi}=0^+$ eigenstate  for $R_c=10$, 12, 14, and 16~a.u., giving values of $^1Z_\text{eff}$ for four different $K$. These values depend on the maximum numbers of partial waves $\ell_{\rm max}$ and radial states per partial wave $n_{\rm max}$ included in Eq.~(\ref{eqn:ps}). We extrapolate in $\ell_{\rm max}$ as ${^1Z}_{\rm eff}(\ell_{\rm max},n_{\rm max}) =  {^1Z}_{\rm eff}(\infty,n_{\rm max}) + A(\ell_{\rm max}+1/2)^{-2}$ and subsequently  in $n_{\rm max}$ as ${^1Z}_{\rm eff}(\infty,n_{\rm max}) = {^1Z}_{\rm eff} + \alpha n^{\beta}_{\rm max},$ where we typically find $\beta\approx-4$ \footnote{Extrapolation in $\ell_\text{max}$ typically reduces the value of ${^1Z}_{\rm eff}$ by ${\sim}20$\%, while extrapolation in $n_\text{max}$ is much less important, changing ${^1Z}_{\rm eff}$ by ${\lesssim}1\%$.}.
The Ps wave function is normalized to the center-of-mass plane wave by comparing the center-of-mass density away from the atom with $\sin^2(KR+\delta_0)/K^2R^2$ (see Ref.~\cite{Swann:thesis} for  details).
Finally, we fit the four values to the effective-range form ${{^1Z}_{\rm eff}(K)\simeq {^1Z}_{\rm eff}(0) + CK^2}$
to deduce ${^1Z}_{\rm eff}(0)$. 
The results are shown in Table~\ref{tab:zeff}. 
Neglecting the enhancement factors, we find good agreement with the previous best zeroth-order results.
Including the enhancement produces near-perfect agreement with 
experimental values for room-temperature Ps.

\textit{Summary.---}The MBT of Ps interactions with atoms was presented and applied to calculate scattering cross sections and pickoff annihilation rates in He and Ne. 
The calculations show that the net effect of the dispersion interaction (electron and positron polarization of the atom and screening of the electron-positron Coulomb interaction by atomic electrons) is relatively small, and close to that described by a model van der Waals potential with a short-range cutoff. The MBT gives pickoff annihilation rates in excellent agreement with experiment. 

\begin{acknowledgments}
\emph{Acknowledgments.---}DGG was supported by the EPSRC UK, grant EP/N007948/1. ARS was supported by the Department for Employment and Leaning, Northern Ireland, UK, and
is supported by the EPSRC UK, grant EP/R006431/1.
\end{acknowledgments}


\begin{thebibliography}{56}%
\makeatletter
\providecommand \@ifxundefined [1]{%
 \@ifx{#1\undefined}
}%
\providecommand \@ifnum [1]{%
 \ifnum #1\expandafter \@firstoftwo
 \else \expandafter \@secondoftwo
 \fi
}%
\providecommand \@ifx [1]{%
 \ifx #1\expandafter \@firstoftwo
 \else \expandafter \@secondoftwo
 \fi
}%
\providecommand \natexlab [1]{#1}%
\providecommand \enquote  [1]{``#1''}%
\providecommand \bibnamefont  [1]{#1}%
\providecommand \bibfnamefont [1]{#1}%
\providecommand \citenamefont [1]{#1}%
\providecommand \href@noop [0]{\@secondoftwo}%
\providecommand \href [0]{\begingroup \@sanitize@url \@href}%
\providecommand \@href[1]{\@@startlink{#1}\@@href}%
\providecommand \@@href[1]{\endgroup#1\@@endlink}%
\providecommand \@sanitize@url [0]{\catcode `\\12\catcode `\$12\catcode
  `\&12\catcode `\#12\catcode `\^12\catcode `\_12\catcode `\%12\relax}%
\providecommand \@@startlink[1]{}%
\providecommand \@@endlink[0]{}%
\providecommand \url  [0]{\begingroup\@sanitize@url \@url }%
\providecommand \@url [1]{\endgroup\@href {#1}{\urlprefix }}%
\providecommand \urlprefix  [0]{URL }%
\providecommand \Eprint [0]{\href }%
\providecommand \doibase [0]{http://dx.doi.org/}%
\providecommand \selectlanguage [0]{\@gobble}%
\providecommand \bibinfo  [0]{\@secondoftwo}%
\providecommand \bibfield  [0]{\@secondoftwo}%
\providecommand \translation [1]{[#1]}%
\providecommand \BibitemOpen [0]{}%
\providecommand \bibitemStop [0]{}%
\providecommand \bibitemNoStop [0]{.\EOS\space}%
\providecommand \EOS [0]{\spacefactor3000\relax}%
\providecommand \BibitemShut  [1]{\csname bibitem#1\endcsname}%
\let\auto@bib@innerbib\@empty
\bibitem [{\citenamefont {Karshenboim}(2005)}]{Karshenboim2005}%
  \BibitemOpen
  \bibfield  {author} {\bibinfo {author} {\bibfnamefont {S.~G.}\ \bibnamefont
  {Karshenboim}},\ }\href {\doibase 10.1016/j.physrep.2005.08.008} {\bibfield
  {journal} {\bibinfo  {journal} {Phys. Rep.}\ }\textbf {\bibinfo {volume}
  {422}},\ \bibinfo {pages} {1} (\bibinfo {year} {2005})}\BibitemShut {NoStop}%
\bibitem [{\citenamefont {Churazov}\ \emph {et~al.}(2011)\citenamefont
  {Churazov}, \citenamefont {Sazonov}, \citenamefont {Tsygankov}, \citenamefont
  {Sunyaev},\ and\ \citenamefont {Varshalovich}}]{MNR:MNR17804}%
  \BibitemOpen
  \bibfield  {author} {\bibinfo {author} {\bibfnamefont {E.}~\bibnamefont
  {Churazov}}, \bibinfo {author} {\bibfnamefont {S.}~\bibnamefont {Sazonov}},
  \bibinfo {author} {\bibfnamefont {S.}~\bibnamefont {Tsygankov}}, \bibinfo
  {author} {\bibfnamefont {R.}~\bibnamefont {Sunyaev}}, \ and\ \bibinfo
  {author} {\bibfnamefont {D.}~\bibnamefont {Varshalovich}},\ }\href {\doibase
  10.1111/j.1365-2966.2010.17804.x} {\bibfield  {journal} {\bibinfo  {journal}
  {Mon. Not. R. Astron. Soc.}\ }\textbf {\bibinfo {volume} {411}},\ \bibinfo
  {pages} {1727} (\bibinfo {year} {2011})}\BibitemShut {NoStop}%
\bibitem [{\citenamefont {Gidley}\ \emph {et~al.}(2006)\citenamefont {Gidley},
  \citenamefont {Peng},\ and\ \citenamefont {Vallery}}]{Gidley06}%
  \BibitemOpen
  \bibfield  {author} {\bibinfo {author} {\bibfnamefont {D.~W.}\ \bibnamefont
  {Gidley}}, \bibinfo {author} {\bibfnamefont {H.}~\bibnamefont {Peng}}, \ and\
  \bibinfo {author} {\bibfnamefont {R.~S.}\ \bibnamefont {Vallery}},\ }\href
  {\doibase 10.1146/annurev.matsci.36.111904.135144} {\bibfield  {journal}
  {\bibinfo  {journal} {Riv. Nuovo. Cimento}\ }\textbf {\bibinfo {volume}
  {36}},\ \bibinfo {pages} {49} (\bibinfo {year} {2006})}\BibitemShut {NoStop}%
\bibitem [{\citenamefont {Kellerbauer}\ \emph {et~al.}(2008)\citenamefont
  {Kellerbauer}, \citenamefont {Amoretti}, \citenamefont {Belov}, \citenamefont
  {Bonomi}, \citenamefont {Boscolo}, \citenamefont {Brusa}, \citenamefont
  {B\"uchner}, \citenamefont {Byakov}, \citenamefont {Cabaret}, \citenamefont
  {Canali}, \citenamefont {Carraro}, \citenamefont {Castelli}, \citenamefont
  {Cialdi}, \citenamefont {de~Combarieu}, \citenamefont {Comparat},
  \citenamefont {Consolati}, \citenamefont {Djourelov}, \citenamefont {Doser},
  \citenamefont {Drobychev}, \citenamefont {Dupasquier}, \citenamefont
  {Ferrari}, \citenamefont {Forget}, \citenamefont {Formaro}, \citenamefont
  {Gervasini}, \citenamefont {Giammarchi}, \citenamefont {Gninenko},
  \citenamefont {Gribakin}, \citenamefont {Hogan}, \citenamefont {Jacquey},
  \citenamefont {Lagomarsino}, \citenamefont {Manuzio}, \citenamefont
  {Mariazzi}, \citenamefont {Matveev}, \citenamefont {Meier}, \citenamefont
  {Merkt}, \citenamefont {Nedelec}, \citenamefont {Oberthaler}, \citenamefont
  {Pari}, \citenamefont {Prevedelli}, \citenamefont {Quasso}, \citenamefont
  {Rotondi}, \citenamefont {Sillou}, \citenamefont {Stepanov}, \citenamefont
  {Stroke}, \citenamefont {Testera}, \citenamefont {Tino}, \citenamefont
  {Tr\'enec}, \citenamefont {Vairo}, \citenamefont {Vigu\'e}, \citenamefont
  {Walters}, \citenamefont {Warring}, \citenamefont {Zavatarelli},\ and\
  \citenamefont {Zvezhinskij}}]{Kellerbauer08}%
  \BibitemOpen
  \bibfield  {author} {\bibinfo {author} {\bibfnamefont {A.}~\bibnamefont
  {Kellerbauer}}, \bibinfo {author} {\bibfnamefont {M.}~\bibnamefont
  {Amoretti}}, \bibinfo {author} {\bibfnamefont {A.}~\bibnamefont {Belov}},
  \bibinfo {author} {\bibfnamefont {G.}~\bibnamefont {Bonomi}}, \bibinfo
  {author} {\bibfnamefont {I.}~\bibnamefont {Boscolo}}, \bibinfo {author}
  {\bibfnamefont {R.}~\bibnamefont {Brusa}}, \bibinfo {author} {\bibfnamefont
  {M.}~\bibnamefont {B\"uchner}}, \bibinfo {author} {\bibfnamefont
  {V.}~\bibnamefont {Byakov}}, \bibinfo {author} {\bibfnamefont
  {L.}~\bibnamefont {Cabaret}}, \bibinfo {author} {\bibfnamefont
  {C.}~\bibnamefont {Canali}}, \bibinfo {author} {\bibfnamefont
  {C.}~\bibnamefont {Carraro}}, \bibinfo {author} {\bibfnamefont
  {F.}~\bibnamefont {Castelli}}, \bibinfo {author} {\bibfnamefont
  {S.}~\bibnamefont {Cialdi}}, \bibinfo {author} {\bibfnamefont
  {M.}~\bibnamefont {de~Combarieu}}, \bibinfo {author} {\bibfnamefont
  {D.}~\bibnamefont {Comparat}}, \bibinfo {author} {\bibfnamefont
  {G.}~\bibnamefont {Consolati}}, \bibinfo {author} {\bibfnamefont
  {N.}~\bibnamefont {Djourelov}}, \bibinfo {author} {\bibfnamefont
  {M.}~\bibnamefont {Doser}}, \bibinfo {author} {\bibfnamefont
  {G.}~\bibnamefont {Drobychev}}, \bibinfo {author} {\bibfnamefont
  {A.}~\bibnamefont {Dupasquier}}, \bibinfo {author} {\bibfnamefont
  {G.}~\bibnamefont {Ferrari}}, \bibinfo {author} {\bibfnamefont
  {P.}~\bibnamefont {Forget}}, \bibinfo {author} {\bibfnamefont
  {L.}~\bibnamefont {Formaro}}, \bibinfo {author} {\bibfnamefont
  {A.}~\bibnamefont {Gervasini}}, \bibinfo {author} {\bibfnamefont
  {M.}~\bibnamefont {Giammarchi}}, \bibinfo {author} {\bibfnamefont
  {S.}~\bibnamefont {Gninenko}}, \bibinfo {author} {\bibfnamefont
  {G.}~\bibnamefont {Gribakin}}, \bibinfo {author} {\bibfnamefont
  {S.}~\bibnamefont {Hogan}}, \bibinfo {author} {\bibfnamefont
  {M.}~\bibnamefont {Jacquey}}, \bibinfo {author} {\bibfnamefont
  {V.}~\bibnamefont {Lagomarsino}}, \bibinfo {author} {\bibfnamefont
  {G.}~\bibnamefont {Manuzio}}, \bibinfo {author} {\bibfnamefont
  {S.}~\bibnamefont {Mariazzi}}, \bibinfo {author} {\bibfnamefont
  {V.}~\bibnamefont {Matveev}}, \bibinfo {author} {\bibfnamefont
  {J.}~\bibnamefont {Meier}}, \bibinfo {author} {\bibfnamefont
  {F.}~\bibnamefont {Merkt}}, \bibinfo {author} {\bibfnamefont
  {P.}~\bibnamefont {Nedelec}}, \bibinfo {author} {\bibfnamefont
  {M.}~\bibnamefont {Oberthaler}}, \bibinfo {author} {\bibfnamefont
  {P.}~\bibnamefont {Pari}}, \bibinfo {author} {\bibfnamefont {M.}~\bibnamefont
  {Prevedelli}}, \bibinfo {author} {\bibfnamefont {F.}~\bibnamefont {Quasso}},
  \bibinfo {author} {\bibfnamefont {A.}~\bibnamefont {Rotondi}}, \bibinfo
  {author} {\bibfnamefont {D.}~\bibnamefont {Sillou}}, \bibinfo {author}
  {\bibfnamefont {S.}~\bibnamefont {Stepanov}}, \bibinfo {author}
  {\bibfnamefont {H.}~\bibnamefont {Stroke}}, \bibinfo {author} {\bibfnamefont
  {G.}~\bibnamefont {Testera}}, \bibinfo {author} {\bibfnamefont
  {G.}~\bibnamefont {Tino}}, \bibinfo {author} {\bibfnamefont {G.}~\bibnamefont
  {Tr\'enec}}, \bibinfo {author} {\bibfnamefont {A.}~\bibnamefont {Vairo}},
  \bibinfo {author} {\bibfnamefont {J.}~\bibnamefont {Vigu\'e}}, \bibinfo
  {author} {\bibfnamefont {H.}~\bibnamefont {Walters}}, \bibinfo {author}
  {\bibfnamefont {U.}~\bibnamefont {Warring}}, \bibinfo {author} {\bibfnamefont
  {S.}~\bibnamefont {Zavatarelli}}, \ and\ \bibinfo {author} {\bibfnamefont
  {D.}~\bibnamefont {Zvezhinskij}},\ }\href {\doibase
  10.1016/j.nimb.2007.12.010} {\bibfield  {journal} {\bibinfo  {journal} {Nucl.
  Instrum. Methods B}\ }\textbf {\bibinfo {volume} {266}},\ \bibinfo {pages}
  {351} (\bibinfo {year} {2008})}\BibitemShut {NoStop}%
\bibitem [{\citenamefont {Cassidy}\ \emph {et~al.}(2011)\citenamefont
  {Cassidy}, \citenamefont {Hisakado}, \citenamefont {Tom},\ and\ \citenamefont
  {{Mills, Jr.}}}]{PhysRevLett.106.173401}%
  \BibitemOpen
  \bibfield  {author} {\bibinfo {author} {\bibfnamefont {D.~B.}\ \bibnamefont
  {Cassidy}}, \bibinfo {author} {\bibfnamefont {T.~H.}\ \bibnamefont
  {Hisakado}}, \bibinfo {author} {\bibfnamefont {H.~W.~K.}\ \bibnamefont
  {Tom}}, \ and\ \bibinfo {author} {\bibfnamefont {A.~P.}\ \bibnamefont
  {{Mills, Jr.}}},\ }\href {\doibase 10.1103/PhysRevLett.106.173401} {\bibfield
   {journal} {\bibinfo  {journal} {Phys. Rev. Lett.}\ }\textbf {\bibinfo
  {volume} {106}},\ \bibinfo {pages} {173401} (\bibinfo {year}
  {2011})}\BibitemShut {NoStop}%
\bibitem [{\citenamefont {Brawley}\ \emph {et~al.}(2015)\citenamefont
  {Brawley}, \citenamefont {Fayer}, \citenamefont {Shipman},\ and\
  \citenamefont {Laricchia}}]{Brawley:2015}%
  \BibitemOpen
  \bibfield  {author} {\bibinfo {author} {\bibfnamefont {S.~J.}\ \bibnamefont
  {Brawley}}, \bibinfo {author} {\bibfnamefont {S.~E.}\ \bibnamefont {Fayer}},
  \bibinfo {author} {\bibfnamefont {M.}~\bibnamefont {Shipman}}, \ and\
  \bibinfo {author} {\bibfnamefont {G.}~\bibnamefont {Laricchia}},\ }\href
  {\doibase 10.1103/PhysRevLett.115.223201} {\bibfield  {journal} {\bibinfo
  {journal} {Phys. Rev. Lett.}\ }\textbf {\bibinfo {volume} {115}},\ \bibinfo
  {pages} {223201} (\bibinfo {year} {2015})}\BibitemShut {NoStop}%
\bibitem [{\citenamefont {Swann}\ and\ \citenamefont
  {Gribakin}(2018)}]{Swann:2018:vdw}%
  \BibitemOpen
  \bibfield  {author} {\bibinfo {author} {\bibfnamefont {A.~R.}\ \bibnamefont
  {Swann}}\ and\ \bibinfo {author} {\bibfnamefont {G.~F.}\ \bibnamefont
  {Gribakin}},\ }\href {\doibase 10.1103/PhysRevA.97.012706} {\bibfield
  {journal} {\bibinfo  {journal} {Phys. Rev. A}\ }\textbf {\bibinfo {volume}
  {97}},\ \bibinfo {pages} {012706} (\bibinfo {year} {2018})}\BibitemShut
  {NoStop}%
\bibitem [{\citenamefont {Fraser}\ and\ \citenamefont
  {Kraidy}(1966)}]{Fraser:1966}%
  \BibitemOpen
  \bibfield  {author} {\bibinfo {author} {\bibfnamefont {P.~A.}\ \bibnamefont
  {Fraser}}\ and\ \bibinfo {author} {\bibfnamefont {M.}~\bibnamefont
  {Kraidy}},\ }\href {http://stacks.iop.org/0370-1328/89/i=3/a=309} {\bibfield
  {journal} {\bibinfo  {journal} {Proc. Phys. Soc.}\ }\textbf {\bibinfo
  {volume} {89}},\ \bibinfo {pages} {533} (\bibinfo {year} {1966})}\BibitemShut
  {NoStop}%
\bibitem [{\citenamefont {Fraser}(1968)}]{Fraser:1968}%
  \BibitemOpen
  \bibfield  {author} {\bibinfo {author} {\bibfnamefont {P.~A.}\ \bibnamefont
  {Fraser}},\ }\href {\doibase 10.1088/0022-3700/1/5/135} {\bibfield  {journal}
  {\bibinfo  {journal} {J. Phys. B}\ }\textbf {\bibinfo {volume} {1}},\
  \bibinfo {pages} {1006} (\bibinfo {year} {1968})}\BibitemShut {NoStop}%
\bibitem [{\citenamefont {Barker}\ and\ \citenamefont
  {Bransden}(1968)}]{Barker:1968}%
  \BibitemOpen
  \bibfield  {author} {\bibinfo {author} {\bibfnamefont {M.~I.}\ \bibnamefont
  {Barker}}\ and\ \bibinfo {author} {\bibfnamefont {B.~H.}\ \bibnamefont
  {Bransden}},\ }\href {\doibase 10.1088/0022-3700/1/6/314} {\bibfield
  {journal} {\bibinfo  {journal} {J. Phys. B}\ }\textbf {\bibinfo {volume}
  {1}},\ \bibinfo {pages} {1109} (\bibinfo {year} {1968})}\BibitemShut
  {NoStop}%
\bibitem [{\citenamefont {Drachman}\ and\ \citenamefont
  {Houston}(1970)}]{Drachman:1970}%
  \BibitemOpen
  \bibfield  {author} {\bibinfo {author} {\bibfnamefont {R.~J.}\ \bibnamefont
  {Drachman}}\ and\ \bibinfo {author} {\bibfnamefont {S.~K.}\ \bibnamefont
  {Houston}},\ }\href {\doibase 10.1088/0022-3700/3/12/010} {\bibfield
  {journal} {\bibinfo  {journal} {J. Phys. B}\ }\textbf {\bibinfo {volume}
  {3}},\ \bibinfo {pages} {1657} (\bibinfo {year} {1970})}\BibitemShut
  {NoStop}%
\bibitem [{\citenamefont {Biswas}\ and\ \citenamefont
  {Adhikari}(2000)}]{Biswas:2000}%
  \BibitemOpen
  \bibfield  {author} {\bibinfo {author} {\bibfnamefont {P.~K.}\ \bibnamefont
  {Biswas}}\ and\ \bibinfo {author} {\bibfnamefont {S.~K.}\ \bibnamefont
  {Adhikari}},\ }\href {\doibase 10.1016/S0009-2614(99)01354-8} {\bibfield
  {journal} {\bibinfo  {journal} {Chem. Phys. Lett.}\ }\textbf {\bibinfo
  {volume} {317}},\ \bibinfo {pages} {129} (\bibinfo {year}
  {2000})}\BibitemShut {NoStop}%
\bibitem [{\citenamefont {Mitroy}\ and\ \citenamefont
  {Ivanov}(2001)}]{Mitroy:2001}%
  \BibitemOpen
  \bibfield  {author} {\bibinfo {author} {\bibfnamefont {J.}~\bibnamefont
  {Mitroy}}\ and\ \bibinfo {author} {\bibfnamefont {I.~A.}\ \bibnamefont
  {Ivanov}},\ }\href {\doibase 10.1103/PhysRevA.65.012509} {\bibfield
  {journal} {\bibinfo  {journal} {Phys. Rev. A}\ }\textbf {\bibinfo {volume}
  {65}},\ \bibinfo {pages} {012509} (\bibinfo {year} {2001})}\BibitemShut
  {NoStop}%
\bibitem [{\citenamefont {Mitroy}\ and\ \citenamefont
  {Bromley}(2003)}]{Mitroy:2003}%
  \BibitemOpen
  \bibfield  {author} {\bibinfo {author} {\bibfnamefont {J.}~\bibnamefont
  {Mitroy}}\ and\ \bibinfo {author} {\bibfnamefont {M.~W.~J.}\ \bibnamefont
  {Bromley}},\ }\href {\doibase 10.1103/PhysRevA.67.034502} {\bibfield
  {journal} {\bibinfo  {journal} {Phys. Rev. A}\ }\textbf {\bibinfo {volume}
  {67}},\ \bibinfo {pages} {034502} (\bibinfo {year} {2003})}\BibitemShut
  {NoStop}%
\bibitem [{\citenamefont {Charlton}(1985)}]{Charlton:1985}%
  \BibitemOpen
  \bibfield  {author} {\bibinfo {author} {\bibfnamefont {M.}~\bibnamefont
  {Charlton}},\ }\href {http://stacks.iop.org/0034-4885/48/i=6/a=001}
  {\bibfield  {journal} {\bibinfo  {journal} {Rep. Prog. Phys.}\ }\textbf
  {\bibinfo {volume} {48}},\ \bibinfo {pages} {737} (\bibinfo {year}
  {1985})}\BibitemShut {NoStop}%
\bibitem [{\citenamefont {Saito}\ and\ \citenamefont
  {Hyodo}(2006)}]{Saito:2006}%
  \BibitemOpen
  \bibfield  {author} {\bibinfo {author} {\bibfnamefont {H.}~\bibnamefont
  {Saito}}\ and\ \bibinfo {author} {\bibfnamefont {T.}~\bibnamefont {Hyodo}},\
  }\href {\doibase 10.1103/PhysRevLett.97.253402} {\bibfield  {journal}
  {\bibinfo  {journal} {Phys. Rev. Lett.}\ }\textbf {\bibinfo {volume} {97}},\
  \bibinfo {pages} {253402} (\bibinfo {year} {2006})}\BibitemShut {NoStop}%
\bibitem [{\citenamefont {Walters}\ \emph {et~al.}(2004)\citenamefont
  {Walters}, \citenamefont {Yu}, \citenamefont {Sahoo},\ and\ \citenamefont
  {Gilmore}}]{Walters:2004}%
  \BibitemOpen
  \bibfield  {author} {\bibinfo {author} {\bibfnamefont {H.}~\bibnamefont
  {Walters}}, \bibinfo {author} {\bibfnamefont {A.}~\bibnamefont {Yu}},
  \bibinfo {author} {\bibfnamefont {S.}~\bibnamefont {Sahoo}}, \ and\ \bibinfo
  {author} {\bibfnamefont {S.}~\bibnamefont {Gilmore}},\ }\href {\doibase
  http://dx.doi.org/10.1016/j.nimb.2004.03.047} {\bibfield  {journal} {\bibinfo
   {journal} {Nucl. Instrum. and Meth. B}\ }\textbf {\bibinfo {volume} {221}},\
  \bibinfo {pages} {149 } (\bibinfo {year} {2004})}\BibitemShut {NoStop}%
\bibitem [{\citenamefont {Dunlop}\ and\ \citenamefont
  {Gribakin}(2006)}]{Dunlop:2006}%
  \BibitemOpen
  \bibfield  {author} {\bibinfo {author} {\bibfnamefont {L.~J.~M.}\
  \bibnamefont {Dunlop}}\ and\ \bibinfo {author} {\bibfnamefont {G.~F.}\
  \bibnamefont {Gribakin}},\ }\href {\doibase 10.1088/0953-4075/39/7/008}
  {\bibfield  {journal} {\bibinfo  {journal} {J. Phys. B}\ }\textbf {\bibinfo
  {volume} {39}},\ \bibinfo {pages} {1647} (\bibinfo {year}
  {2006})}\BibitemShut {NoStop}%
\bibitem [{\citenamefont {Green}\ \emph {et~al.}(2014)\citenamefont {Green},
  \citenamefont {Ludlow},\ and\ \citenamefont {Gribakin}}]{DGG_posnobles}%
  \BibitemOpen
  \bibfield  {author} {\bibinfo {author} {\bibfnamefont {D.~G.}\ \bibnamefont
  {Green}}, \bibinfo {author} {\bibfnamefont {J.~A.}\ \bibnamefont {Ludlow}}, \
  and\ \bibinfo {author} {\bibfnamefont {G.~F.}\ \bibnamefont {Gribakin}},\
  }\href {\doibase 10.1103/PhysRevA.90.032712} {\bibfield  {journal} {\bibinfo
  {journal} {Phys. Rev. A}\ }\textbf {\bibinfo {volume} {90}},\ \bibinfo
  {pages} {032712} (\bibinfo {year} {2014})}\BibitemShut {NoStop}%
\bibitem [{\citenamefont {Green}\ and\ \citenamefont
  {Gribakin}(2015)}]{DGG:2015:core}%
  \BibitemOpen
  \bibfield  {author} {\bibinfo {author} {\bibfnamefont {D.~G.}\ \bibnamefont
  {Green}}\ and\ \bibinfo {author} {\bibfnamefont {G.~F.}\ \bibnamefont
  {Gribakin}},\ }\href {\doibase 10.1103/PhysRevLett.114.093201} {\bibfield
  {journal} {\bibinfo  {journal} {Phys. Rev. Lett.}\ }\textbf {\bibinfo
  {volume} {114}},\ \bibinfo {pages} {093201} (\bibinfo {year}
  {2015})}\BibitemShut {NoStop}%
\bibitem [{\citenamefont {Green}\ and\ \citenamefont
  {Gribakin}(2013)}]{DGG_hlike}%
  \BibitemOpen
  \bibfield  {author} {\bibinfo {author} {\bibfnamefont {D.~G.}\ \bibnamefont
  {Green}}\ and\ \bibinfo {author} {\bibfnamefont {G.~F.}\ \bibnamefont
  {Gribakin}},\ }\href {\doibase 10.1103/PhysRevA.88.032708} {\bibfield
  {journal} {\bibinfo  {journal} {Phys. Rev. A}\ }\textbf {\bibinfo {volume}
  {88}},\ \bibinfo {pages} {032708} (\bibinfo {year} {2013})}\BibitemShut
  {NoStop}%
\bibitem [{Note1()}]{Note1}%
  \BibitemOpen
  \bibinfo {note} {The only exception is a calculation for Ps-H$_2$ that uses
  explicitly correlated Gaussians \cite {Zhang:2018}.}\BibitemShut {Stop}%
\bibitem [{\citenamefont {Kelly}(1967)}]{Kelly:MBT:elcAtom}%
  \BibitemOpen
  \bibfield  {author} {\bibinfo {author} {\bibfnamefont {H.~P.}\ \bibnamefont
  {Kelly}},\ }\href {\doibase 10.1103/PhysRev.160.44} {\bibfield  {journal}
  {\bibinfo  {journal} {Phys. Rev.}\ }\textbf {\bibinfo {volume} {160}},\
  \bibinfo {pages} {44} (\bibinfo {year} {1967})}\BibitemShut {NoStop}%
\bibitem [{\citenamefont {Amusia}\ \emph {et~al.}(1974)\citenamefont {Amusia},
  \citenamefont {Cherepkov}, \citenamefont {Chernysheva},\ and\ \citenamefont
  {Shapiro}}]{Amusia:elc:MBT:Ar1}%
  \BibitemOpen
  \bibfield  {author} {\bibinfo {author} {\bibfnamefont {M.}~\bibnamefont
  {Amusia}}, \bibinfo {author} {\bibfnamefont {N.}~\bibnamefont {Cherepkov}},
  \bibinfo {author} {\bibfnamefont {L.}~\bibnamefont {Chernysheva}}, \ and\
  \bibinfo {author} {\bibfnamefont {S.}~\bibnamefont {Shapiro}},\ }\href
  {\doibase 10.1016/0375-9601(74)90929-3} {\bibfield  {journal} {\bibinfo
  {journal} {Phys. Lett. A}\ }\textbf {\bibinfo {volume} {46}},\ \bibinfo
  {pages} {387 } (\bibinfo {year} {1974})}\BibitemShut {NoStop}%
\bibitem [{\citenamefont {Amusia}\ \emph {et~al.}(1975)\citenamefont {Amusia},
  \citenamefont {Cherepkov}, \citenamefont {Tancic}, \citenamefont {Shapiro},\
  and\ \citenamefont {Chernysheva}}]{Amusia:JETP:1975}%
  \BibitemOpen
  \bibfield  {author} {\bibinfo {author} {\bibfnamefont {M.~Y.}\ \bibnamefont
  {Amusia}}, \bibinfo {author} {\bibfnamefont {N.~A.}\ \bibnamefont
  {Cherepkov}}, \bibinfo {author} {\bibfnamefont {A.}~\bibnamefont {Tancic}},
  \bibinfo {author} {\bibfnamefont {S.~G.}\ \bibnamefont {Shapiro}}, \ and\
  \bibinfo {author} {\bibfnamefont {L.}~\bibnamefont {Chernysheva}},\
  }\href@noop {} {\bibfield  {journal} {\bibinfo  {journal} {Zh. Eksp. Teor.
  Phys.}\ }\textbf {\bibinfo {volume} {68}},\ \bibinfo {pages} {2023} (\bibinfo
  {year} {1975})},\ \bibinfo {note} {[Sov. Phys. JETP {\bf 41}, 1012
  (1975)]}\BibitemShut {NoStop}%
\bibitem [{\citenamefont {Amusia}\ \emph {et~al.}(1982)\citenamefont {Amusia},
  \citenamefont {Cherepkov}, \citenamefont {Chernysheva}, \citenamefont
  {Davidovi\ifmmode~\acute{c}\else \'{c}\fi{}},\ and\ \citenamefont
  {Radojevi\ifmmode~\acute{c}\else \'{c}\fi{}}}]{Amusia:elc:MBT:Ar2}%
  \BibitemOpen
  \bibfield  {author} {\bibinfo {author} {\bibfnamefont {M.~Y.}\ \bibnamefont
  {Amusia}}, \bibinfo {author} {\bibfnamefont {N.~A.}\ \bibnamefont
  {Cherepkov}}, \bibinfo {author} {\bibfnamefont {L.~V.}\ \bibnamefont
  {Chernysheva}}, \bibinfo {author} {\bibfnamefont {D.~M.}\ \bibnamefont
  {Davidovi\ifmmode~\acute{c}\else \'{c}\fi{}}}, \ and\ \bibinfo {author}
  {\bibfnamefont {V.}~\bibnamefont {Radojevi\ifmmode~\acute{c}\else
  \'{c}\fi{}}},\ }\href {\doibase 10.1103/PhysRevA.25.219} {\bibfield
  {journal} {\bibinfo  {journal} {Phys. Rev. A}\ }\textbf {\bibinfo {volume}
  {25}},\ \bibinfo {pages} {219} (\bibinfo {year} {1982})}\BibitemShut
  {NoStop}%
\bibitem [{\citenamefont {Johnson}\ and\ \citenamefont
  {Guet}(1994)}]{Johnson:MBT:elc:Xe}%
  \BibitemOpen
  \bibfield  {author} {\bibinfo {author} {\bibfnamefont {W.~R.}\ \bibnamefont
  {Johnson}}\ and\ \bibinfo {author} {\bibfnamefont {C.}~\bibnamefont {Guet}},\
  }\href {\doibase 10.1103/PhysRevA.49.1041} {\bibfield  {journal} {\bibinfo
  {journal} {Phys. Rev. A}\ }\textbf {\bibinfo {volume} {49}},\ \bibinfo
  {pages} {1041} (\bibinfo {year} {1994})}\BibitemShut {NoStop}%
\bibitem [{\citenamefont {Cheng}\ \emph {et~al.}(2014)\citenamefont {Cheng},
  \citenamefont {Tang}, \citenamefont {Mitroy},\ and\ \citenamefont
  {Safronova}}]{Safronova:MBT:elcAtom}%
  \BibitemOpen
  \bibfield  {author} {\bibinfo {author} {\bibfnamefont {Y.}~\bibnamefont
  {Cheng}}, \bibinfo {author} {\bibfnamefont {L.~Y.}\ \bibnamefont {Tang}},
  \bibinfo {author} {\bibfnamefont {J.}~\bibnamefont {Mitroy}}, \ and\ \bibinfo
  {author} {\bibfnamefont {M.~S.}\ \bibnamefont {Safronova}},\ }\href {\doibase
  10.1103/PhysRevA.89.012701} {\bibfield  {journal} {\bibinfo  {journal} {Phys.
  Rev. A}\ }\textbf {\bibinfo {volume} {89}},\ \bibinfo {pages} {012701}
  (\bibinfo {year} {2014})}\BibitemShut {NoStop}%
\bibitem [{\citenamefont {Amusia}\ \emph {et~al.}(1976)\citenamefont {Amusia},
  \citenamefont {Cherepkov}, \citenamefont {Chernysheva},\ and\ \citenamefont
  {Shapiro}}]{Amusia:Pos:MBT:He}%
  \BibitemOpen
  \bibfield  {author} {\bibinfo {author} {\bibfnamefont {M.~Y.}\ \bibnamefont
  {Amusia}}, \bibinfo {author} {\bibfnamefont {N.~A.}\ \bibnamefont
  {Cherepkov}}, \bibinfo {author} {\bibfnamefont {L.~V.}\ \bibnamefont
  {Chernysheva}}, \ and\ \bibinfo {author} {\bibfnamefont {S.~G.}\ \bibnamefont
  {Shapiro}},\ }\href {http://stacks.iop.org/0022-3700/9/i=17/a=005} {\bibfield
   {journal} {\bibinfo  {journal} {J. Phys. B}\ }\textbf {\bibinfo {volume}
  {9}},\ \bibinfo {pages} {L531} (\bibinfo {year} {1976})}\BibitemShut
  {NoStop}%
\bibitem [{\citenamefont {Dzuba}\ \emph {et~al.}(1993)\citenamefont {Dzuba},
  \citenamefont {Flambaum}, \citenamefont {King}, \citenamefont {Miller},\ and\
  \citenamefont {Sushkov}}]{PhysScripta.46.248}%
  \BibitemOpen
  \bibfield  {author} {\bibinfo {author} {\bibfnamefont {V.~A.}\ \bibnamefont
  {Dzuba}}, \bibinfo {author} {\bibfnamefont {V.~V.}\ \bibnamefont {Flambaum}},
  \bibinfo {author} {\bibfnamefont {W.~A.}\ \bibnamefont {King}}, \bibinfo
  {author} {\bibfnamefont {B.~N.}\ \bibnamefont {Miller}}, \ and\ \bibinfo
  {author} {\bibfnamefont {O.~P.}\ \bibnamefont {Sushkov}},\ }\href {\doibase
  10.1088/0031-8949/1993/T46/039} {\bibfield  {journal} {\bibinfo  {journal}
  {Phys. Scripta}\ }\textbf {\bibinfo {volume} {T46}},\ \bibinfo {pages} {248}
  (\bibinfo {year} {1993})}\BibitemShut {NoStop}%
\bibitem [{\citenamefont {Gribakin}\ and\ \citenamefont
  {Ludlow}(2004)}]{Gribakin:2004}%
  \BibitemOpen
  \bibfield  {author} {\bibinfo {author} {\bibfnamefont {G.~F.}\ \bibnamefont
  {Gribakin}}\ and\ \bibinfo {author} {\bibfnamefont {J.}~\bibnamefont
  {Ludlow}},\ }\href {\doibase 10.1103/PhysRevA.70.032720} {\bibfield
  {journal} {\bibinfo  {journal} {Phys. Rev. A}\ }\textbf {\bibinfo {volume}
  {70}},\ \bibinfo {pages} {032720} (\bibinfo {year} {2004})}\BibitemShut
  {NoStop}%
\bibitem [{\citenamefont {Green}(2017{\natexlab{a}})}]{DGG_cool}%
  \BibitemOpen
  \bibfield  {author} {\bibinfo {author} {\bibfnamefont {D.~G.}\ \bibnamefont
  {Green}},\ }\href {\doibase 10.1103/PhysRevLett.119.203403} {\bibfield
  {journal} {\bibinfo  {journal} {Phys. Rev. Lett.}\ }\textbf {\bibinfo
  {volume} {119}},\ \bibinfo {pages} {203403} (\bibinfo {year}
  {2017}{\natexlab{a}})}\BibitemShut {NoStop}%
\bibitem [{\citenamefont {Green}(2017{\natexlab{b}})}]{DGG_gamcool}%
  \BibitemOpen
  \bibfield  {author} {\bibinfo {author} {\bibfnamefont {D.~G.}\ \bibnamefont
  {Green}},\ }\href {\doibase 10.1103/PhysRevLett.119.203404} {\bibfield
  {journal} {\bibinfo  {journal} {Phys. Rev. Lett.}\ }\textbf {\bibinfo
  {volume} {119}},\ \bibinfo {pages} {203404} (\bibinfo {year}
  {2017}{\natexlab{b}})}\BibitemShut {NoStop}%
\bibitem [{\citenamefont {Fetter}\ and\ \citenamefont
  {Walecka}(2003)}]{fetterwalecka}%
  \BibitemOpen
  \bibfield  {author} {\bibinfo {author} {\bibfnamefont {A.~L.}\ \bibnamefont
  {Fetter}}\ and\ \bibinfo {author} {\bibfnamefont {J.~D.}\ \bibnamefont
  {Walecka}},\ }\href@noop {} {\emph {\bibinfo {title} {Quantum theory of
  many-particle systems}}}\ (\bibinfo  {publisher} {Dover, New York},\ \bibinfo
  {year} {2003})\BibitemShut {NoStop}%
\bibitem [{Note2()}]{Note2}%
  \BibitemOpen
  \bibinfo {note} {$\protect \mathaccentV {hat}05E{\Sigma }^{\pm }_{E}$ acts as
  $\protect \mathaccentV {hat}05E{\Sigma }^\pm _E\psi _{\varepsilon }({\protect
  \bf r}) = \DOTSI \intop \ilimits@ \Sigma _E^\pm ({\protect \bf r},{\protect
  \bf r}')\psi _{\varepsilon }({\protect \bf r'})\protect \tmspace +\thinmuskip
  {.1667em} d{\protect \bf r}'$.}\BibitemShut {Stop}%
\bibitem [{Note3()}]{Note3}%
  \BibitemOpen
  \bibinfo {note} {See Ref.~\cite {Safronova:MBT:elcAtom} for higher-order
  calculations.}\BibitemShut {Stop}%
\bibitem [{\citenamefont {Dzuba}\ and\ \citenamefont
  {Gribakin}(1994)}]{Dzuba:1994}%
  \BibitemOpen
  \bibfield  {author} {\bibinfo {author} {\bibfnamefont {V.~A.}\ \bibnamefont
  {Dzuba}}\ and\ \bibinfo {author} {\bibfnamefont {G.~F.}\ \bibnamefont
  {Gribakin}},\ }\href {\doibase 10.1103/PhysRevA.49.2483} {\bibfield
  {journal} {\bibinfo  {journal} {Phys. Rev. A}\ }\textbf {\bibinfo {volume}
  {49}},\ \bibinfo {pages} {2483} (\bibinfo {year} {1994})}\BibitemShut
  {NoStop}%
\bibitem [{Note4()}]{Note4}%
  \BibitemOpen
  \bibinfo {note} {There is a similarity between our approach and the
  combination of MBT with the configuration-interaction method for open-shell
  atoms \cite {Dzuba:MBTCI}}\BibitemShut {NoStop}%
\bibitem [{\citenamefont {Brown}\ \emph {et~al.}(2017)\citenamefont {Brown},
  \citenamefont {Prigent}, \citenamefont {Swann},\ and\ \citenamefont
  {Gribakin}}]{Brown:2017}%
  \BibitemOpen
  \bibfield  {author} {\bibinfo {author} {\bibfnamefont {R.}~\bibnamefont
  {Brown}}, \bibinfo {author} {\bibfnamefont {Q.}~\bibnamefont {Prigent}},
  \bibinfo {author} {\bibfnamefont {A.~R.}\ \bibnamefont {Swann}}, \ and\
  \bibinfo {author} {\bibfnamefont {G.~F.}\ \bibnamefont {Gribakin}},\ }\href
  {\doibase 10.1103/PhysRevA.95.032705} {\bibfield  {journal} {\bibinfo
  {journal} {Phys. Rev. A}\ }\textbf {\bibinfo {volume} {95}},\ \bibinfo
  {pages} {032705} (\bibinfo {year} {2017})}\BibitemShut {NoStop}%
\bibitem [{\citenamefont {Canter}\ \emph {et~al.}(1975)\citenamefont {Canter},
  \citenamefont {McNutt},\ and\ \citenamefont {Roellig}}]{Canter:1975}%
  \BibitemOpen
  \bibfield  {author} {\bibinfo {author} {\bibfnamefont {K.~F.}\ \bibnamefont
  {Canter}}, \bibinfo {author} {\bibfnamefont {J.~D.}\ \bibnamefont {McNutt}},
  \ and\ \bibinfo {author} {\bibfnamefont {L.~O.}\ \bibnamefont {Roellig}},\
  }\href {\doibase 10.1103/PhysRevA.12.375} {\bibfield  {journal} {\bibinfo
  {journal} {Phys. Rev. A}\ }\textbf {\bibinfo {volume} {12}},\ \bibinfo
  {pages} {375} (\bibinfo {year} {1975})}\BibitemShut {NoStop}%
\bibitem [{\citenamefont {Rytsola}\ \emph {et~al.}(1984)\citenamefont
  {Rytsola}, \citenamefont {Vettenranta},\ and\ \citenamefont
  {Hautojarvi}}]{Rytsola:1984}%
  \BibitemOpen
  \bibfield  {author} {\bibinfo {author} {\bibfnamefont {K.}~\bibnamefont
  {Rytsola}}, \bibinfo {author} {\bibfnamefont {J.}~\bibnamefont
  {Vettenranta}}, \ and\ \bibinfo {author} {\bibfnamefont {P.}~\bibnamefont
  {Hautojarvi}},\ }\href {\doibase 10.1088/0022-3700/17/16/018} {\bibfield
  {journal} {\bibinfo  {journal} {J. Phys. B}\ }\textbf {\bibinfo {volume}
  {17}},\ \bibinfo {pages} {3359} (\bibinfo {year} {1984})}\BibitemShut
  {NoStop}%
\bibitem [{\citenamefont {Coleman}\ \emph {et~al.}(1994)\citenamefont
  {Coleman}, \citenamefont {Rayner}, \citenamefont {Jacobsen}, \citenamefont
  {Charlton},\ and\ \citenamefont {West}}]{Coleman:1994}%
  \BibitemOpen
  \bibfield  {author} {\bibinfo {author} {\bibfnamefont {P.~G.}\ \bibnamefont
  {Coleman}}, \bibinfo {author} {\bibfnamefont {S.}~\bibnamefont {Rayner}},
  \bibinfo {author} {\bibfnamefont {F.~M.}\ \bibnamefont {Jacobsen}}, \bibinfo
  {author} {\bibfnamefont {M.}~\bibnamefont {Charlton}}, \ and\ \bibinfo
  {author} {\bibfnamefont {R.~N.}\ \bibnamefont {West}},\ }\href {\doibase
  10.1088/0953-4075/27/5/017} {\bibfield  {journal} {\bibinfo  {journal} {J.
  Phys. B}\ }\textbf {\bibinfo {volume} {27}},\ \bibinfo {pages} {981}
  (\bibinfo {year} {1994})}\BibitemShut {NoStop}%
\bibitem [{\citenamefont {Skalsey}\ \emph {et~al.}(2003)\citenamefont
  {Skalsey}, \citenamefont {Engbrecht}, \citenamefont {Nakamura}, \citenamefont
  {Vallery},\ and\ \citenamefont {Gidley}}]{Skalsey:2003}%
  \BibitemOpen
  \bibfield  {author} {\bibinfo {author} {\bibfnamefont {M.}~\bibnamefont
  {Skalsey}}, \bibinfo {author} {\bibfnamefont {J.~J.}\ \bibnamefont
  {Engbrecht}}, \bibinfo {author} {\bibfnamefont {C.~M.}\ \bibnamefont
  {Nakamura}}, \bibinfo {author} {\bibfnamefont {R.~S.}\ \bibnamefont
  {Vallery}}, \ and\ \bibinfo {author} {\bibfnamefont {D.~W.}\ \bibnamefont
  {Gidley}},\ }\href {\doibase 10.1103/PhysRevA.67.022504} {\bibfield
  {journal} {\bibinfo  {journal} {Phys. Rev. A}\ }\textbf {\bibinfo {volume}
  {67}},\ \bibinfo {pages} {022504} (\bibinfo {year} {2003})}\BibitemShut
  {NoStop}%
\bibitem [{\citenamefont {Nagashima}\ \emph {et~al.}(1998)\citenamefont
  {Nagashima}, \citenamefont {Hyodo}, \citenamefont {Fujiwara},\ and\
  \citenamefont {Ichimura}}]{Nagashima:1998}%
  \BibitemOpen
  \bibfield  {author} {\bibinfo {author} {\bibfnamefont {Y.}~\bibnamefont
  {Nagashima}}, \bibinfo {author} {\bibfnamefont {T.}~\bibnamefont {Hyodo}},
  \bibinfo {author} {\bibfnamefont {K.}~\bibnamefont {Fujiwara}}, \ and\
  \bibinfo {author} {\bibfnamefont {A.}~\bibnamefont {Ichimura}},\ }\href
  {\doibase 10.1088/0953-4075/31/2/014} {\bibfield  {journal} {\bibinfo
  {journal} {J. Phys. B}\ }\textbf {\bibinfo {volume} {31}},\ \bibinfo {pages}
  {329} (\bibinfo {year} {1998})}\BibitemShut {NoStop}%
\bibitem [{\citenamefont {Saito}\ \emph {et~al.}(2003)\citenamefont {Saito},
  \citenamefont {Nagashima},\ and\ \citenamefont {Hyodo}}]{Saito:2003}%
  \BibitemOpen
  \bibfield  {author} {\bibinfo {author} {\bibfnamefont {F.}~\bibnamefont
  {Saito}}, \bibinfo {author} {\bibfnamefont {Y.}~\bibnamefont {Nagashima}}, \
  and\ \bibinfo {author} {\bibfnamefont {T.}~\bibnamefont {Hyodo}},\ }\href
  {\doibase 10.1088/0953-4075/36/20/011} {\bibfield  {journal} {\bibinfo
  {journal} {J. Phys. B}\ }\textbf {\bibinfo {volume} {36}},\ \bibinfo {pages}
  {4191} (\bibinfo {year} {2003})}\BibitemShut {NoStop}%
\bibitem [{\citenamefont {Engbrecht}\ \emph {et~al.}(2008)\citenamefont
  {Engbrecht}, \citenamefont {Erickson}, \citenamefont {Johnson}, \citenamefont
  {Kolan}, \citenamefont {Legard}, \citenamefont {Lund}, \citenamefont
  {Nyflot},\ and\ \citenamefont {Paulsen}}]{Engbrecht:2008}%
  \BibitemOpen
  \bibfield  {author} {\bibinfo {author} {\bibfnamefont {J.~J.}\ \bibnamefont
  {Engbrecht}}, \bibinfo {author} {\bibfnamefont {M.~J.}\ \bibnamefont
  {Erickson}}, \bibinfo {author} {\bibfnamefont {C.~P.}\ \bibnamefont
  {Johnson}}, \bibinfo {author} {\bibfnamefont {A.~J.}\ \bibnamefont {Kolan}},
  \bibinfo {author} {\bibfnamefont {A.~E.}\ \bibnamefont {Legard}}, \bibinfo
  {author} {\bibfnamefont {S.~P.}\ \bibnamefont {Lund}}, \bibinfo {author}
  {\bibfnamefont {M.~J.}\ \bibnamefont {Nyflot}}, \ and\ \bibinfo {author}
  {\bibfnamefont {J.~D.}\ \bibnamefont {Paulsen}},\ }\href {\doibase
  10.1103/PhysRevA.77.012711} {\bibfield  {journal} {\bibinfo  {journal} {Phys.
  Rev. A}\ }\textbf {\bibinfo {volume} {77}},\ \bibinfo {pages} {012711}
  (\bibinfo {year} {2008})}\BibitemShut {NoStop}%
\bibitem [{Note5()}]{Note5}%
  \BibitemOpen
  \bibinfo {note} {The FT+vdW data corresponds to the cutoff radius
  $R_0=3.0$~a.u. in the model van der Waals potential \cite
  {Swann:2018:vdw}.}\BibitemShut {Stop}%
\bibitem [{Note6()}]{Note6}%
  \BibitemOpen
  \bibinfo {note} {We have checked that a Ps-atom potential of the form
  $U(R)=B/R^{12}-C_6/R^6$ (as suggested in Ref.~\cite {Nieminen:1980}) with
  accurate $C_6$ \cite {Swann:2015} and $B$ chosen to fit the MBT scattering
  length, gives the elastic cross sections that exceed the MBT result by 50\%
  (Ne) to 100\% (He) at $K\sim 1$~a.u., due to overestimated effect of
  repulsion.}\BibitemShut {Stop}%
\bibitem [{\citenamefont {Barker}\ and\ \citenamefont
  {Bransden}(1969)}]{Barker:1969}%
  \BibitemOpen
  \bibfield  {author} {\bibinfo {author} {\bibfnamefont {M.~I.}\ \bibnamefont
  {Barker}}\ and\ \bibinfo {author} {\bibfnamefont {B.~H.}\ \bibnamefont
  {Bransden}},\ }\href {\doibase 10.1088/0022-3700/2/6/515} {\bibfield
  {journal} {\bibinfo  {journal} {J. Phys. B}\ }\textbf {\bibinfo {volume}
  {2}},\ \bibinfo {pages} {730} (\bibinfo {year} {1969})}\BibitemShut {NoStop}%
\bibitem [{\citenamefont {Green}\ and\ \citenamefont
  {Gribakin}(2018)}]{DGG:2017:ef}%
  \BibitemOpen
  \bibfield  {author} {\bibinfo {author} {\bibfnamefont {D.~G.}\ \bibnamefont
  {Green}}\ and\ \bibinfo {author} {\bibfnamefont {G.~F.}\ \bibnamefont
  {Gribakin}},\ }\href@noop {} {\bibfield  {journal} {\bibinfo  {journal}
  {Prog. Theor. Chem. and Phys.}\ } (\bibinfo {year} {2018})},\ \Eprint
  {http://arxiv.org/abs/arXiv:1703.06980} {arXiv:1703.06980} \BibitemShut
  {NoStop}%
\bibitem [{Note7()}]{Note7}%
  \BibitemOpen
  \bibinfo {note} {Extrapolation in $\ell _\protect \text {max}$ typically
  reduces the value of ${^1Z}_{\protect \rm eff}$ by ${\sim }20$\%, while
  extrapolation in $n_\protect \text {max}$ is much less important, changing
  ${^1Z}_{\protect \rm eff}$ by ${\lesssim }1\%$.}\BibitemShut {Stop}%
\bibitem [{\citenamefont {Swann}(2017)}]{Swann:thesis}%
  \BibitemOpen
  \bibfield  {author} {\bibinfo {author} {\bibfnamefont {A.~R.}\ \bibnamefont
  {Swann}},\ }\href@noop {} {Ph.D. thesis},\ \bibinfo  {school} {Queen's
  University Belfast} (\bibinfo {year} {2017})\BibitemShut {NoStop}%
\bibitem [{\citenamefont {Zhang}\ \emph {et~al.}(2018)\citenamefont {Zhang},
  \citenamefont {Wu}, \citenamefont {Qian}, \citenamefont {Gao}, \citenamefont
  {Yang}, \citenamefont {Varga}, \citenamefont {Yan},\ and\ \citenamefont
  {Schwingenschl\"ogl}}]{Zhang:2018}%
  \BibitemOpen
  \bibfield  {author} {\bibinfo {author} {\bibfnamefont {J.-Y.}\ \bibnamefont
  {Zhang}}, \bibinfo {author} {\bibfnamefont {M.-S.}\ \bibnamefont {Wu}},
  \bibinfo {author} {\bibfnamefont {Y.}~\bibnamefont {Qian}}, \bibinfo {author}
  {\bibfnamefont {X.}~\bibnamefont {Gao}}, \bibinfo {author} {\bibfnamefont
  {Y.-J.}\ \bibnamefont {Yang}}, \bibinfo {author} {\bibfnamefont
  {K.}~\bibnamefont {Varga}}, \bibinfo {author} {\bibfnamefont {Z.-C.}\
  \bibnamefont {Yan}}, \ and\ \bibinfo {author} {\bibfnamefont
  {U.}~\bibnamefont {Schwingenschl\"ogl}},\ }\href@noop {} {\enquote {\bibinfo
  {title} {S-wave elastic scattering of $o$-{Ps} from {H}$_2$ at low energy},}\
  } (\bibinfo {year} {2018}),\ \Eprint {http://arxiv.org/abs/arXiv:1803.03026}
  {arXiv:1803.03026} \BibitemShut {NoStop}%
\bibitem [{\citenamefont {Dzuba}\ \emph {et~al.}(1996)\citenamefont {Dzuba},
  \citenamefont {Flambaum},\ and\ \citenamefont {Kozlov}}]{Dzuba:MBTCI}%
  \BibitemOpen
  \bibfield  {author} {\bibinfo {author} {\bibfnamefont {V.~A.}\ \bibnamefont
  {Dzuba}}, \bibinfo {author} {\bibfnamefont {V.~V.}\ \bibnamefont {Flambaum}},
  \ and\ \bibinfo {author} {\bibfnamefont {M.~G.}\ \bibnamefont {Kozlov}},\
  }\href {\doibase 10.1103/PhysRevA.54.3948} {\bibfield  {journal} {\bibinfo
  {journal} {Phys. Rev. A}\ }\textbf {\bibinfo {volume} {54}},\ \bibinfo
  {pages} {3948} (\bibinfo {year} {1996})}\BibitemShut {NoStop}%
\bibitem [{\citenamefont {Nieminen}\ \emph {et~al.}(1980)\citenamefont
  {Nieminen}, \citenamefont {V\"alimaa}, \citenamefont {Manninen},\ and\
  \citenamefont {Hautoj\"arvi}}]{Nieminen:1980}%
  \BibitemOpen
  \bibfield  {author} {\bibinfo {author} {\bibfnamefont {R.~M.}\ \bibnamefont
  {Nieminen}}, \bibinfo {author} {\bibfnamefont {I.}~\bibnamefont {V\"alimaa}},
  \bibinfo {author} {\bibfnamefont {M.}~\bibnamefont {Manninen}}, \ and\
  \bibinfo {author} {\bibfnamefont {P.}~\bibnamefont {Hautoj\"arvi}},\ }\href
  {\doibase 10.1103/PhysRevA.21.1677} {\bibfield  {journal} {\bibinfo
  {journal} {Phys. Rev. A}\ }\textbf {\bibinfo {volume} {21}},\ \bibinfo
  {pages} {1677} (\bibinfo {year} {1980})}\BibitemShut {NoStop}%
\bibitem [{\citenamefont {Swann}\ \emph {et~al.}(2015)\citenamefont {Swann},
  \citenamefont {Ludlow},\ and\ \citenamefont {Gribakin}}]{Swann:2015}%
  \BibitemOpen
  \bibfield  {author} {\bibinfo {author} {\bibfnamefont {A.~R.}\ \bibnamefont
  {Swann}}, \bibinfo {author} {\bibfnamefont {J.~A.}\ \bibnamefont {Ludlow}}, \
  and\ \bibinfo {author} {\bibfnamefont {G.~F.}\ \bibnamefont {Gribakin}},\
  }\href {\doibase 10.1103/PhysRevA.92.012505} {\bibfield  {journal} {\bibinfo
  {journal} {Phys. Rev. A}\ }\textbf {\bibinfo {volume} {92}},\ \bibinfo
  {pages} {012505} (\bibinfo {year} {2015})}\BibitemShut {NoStop}%
\end{thebibliography}

%

\end{document}